\documentclass[10pt,journal,compsoc,twoside]{IEEEtran}

\ifCLASSOPTIONcompsoc
\usepackage[nocompress]{cite}
\else
\usepackage{cite}
\fi

\usepackage[T1]{fontenc}% optional T1 font encoding
\usepackage{amsmath}

\usepackage{amssymb} 	% For the purpose to use symbol like leqslant. 

\usepackage[cmintegrals]{newtxmath}
\usepackage{bm} % optional

\usepackage{graphicx}

\usepackage[utf8]{inputenc}

\usepackage{cite}		% For the purpose of ordered inline citing
\usepackage{amsthm}		% For the purpose of writing definition and/or proof of theorem

\usepackage{comment}
\usepackage{makecell} 	% For the purpose of showing multi-line content in table cell
\usepackage{mathrsfs} 	% For the purpose of using mathscr font (e.g. A, P, etc)
\usepackage{amsfonts} 	% For the purpose of using mathbb font (e.g. R, N, etc)
\usepackage{xcolor}
\usepackage{enumitem}	% For the purpose of managing indent and margin of itemize environment.
\setlist{leftmargin=1.75em}
\usepackage{url}

\usepackage{breakurl}
\usepackage[breaklinks]{hyperref}

\usepackage{longtable, afterpage}    % For the purpose to add long table for surveyed papers
\usepackage{supertabular,booktabs}	 % For the purpose to add long table for surveyed papers (experiment)
\usepackage{multicol, supertabular}  % For the purpose to add long table for surveyed papers
\usepackage{pdflscape}               % For the purpose to add long table for surveyed papers (optional)
\usepackage{array}                   % For the purpose to add long table for surveyed papers
\usepackage{multirow}                % For the purpose to add long table for surveyed papers
\usepackage{booktabs}                % For the purpose to add long table for surveyed papers
\usepackage{tabularx}                % For the purpose to add long table for surveyed papers
\usepackage{color}		 % For the purpose to add color to the text, may remove it for the submission version
\usepackage{datetime}				 % Show version date, to be removed in the final release
\usepackage{subfigure}
\usepackage{shortcuts}
\usepackage{tikz}
\newcommand*\emptycirc{\tikz\draw (0,0) circle (0.85ex);} 

\newcommand*\fullcirc{\tikz\fill (0,0) circle (0.85ex);} 

\newcolumntype{H}{>{\setbox0=\hbox\bgroup}c<{\egroup}@{}}

\newcolumntype{Z}{>{\centering\arraybackslash}X}

\newcounter{magicrownumbers}

\newtheorem{definition}{Definition}

\newcounter{rownumber}
\begin{document}

\title{Adversarial Robustness of\\Deep Neural Networks: A Survey\\from a Formal Verification Perspective}

\begin{comment}
\author{Anonymous Authors}
\end{comment}

\author{Mark~Huasong~Meng,
	Guangdong~Bai,
	Sin~Gee~Teo,
	Zhe~Hou,\\
	Yan~Xiao,
	Yun~Lin,
	Jin~Song~Dong
	% Co~Authors,~\IEEEmembership{Fellow,~OSA,}% <-this % stops a space
	\IEEEcompsocitemizethanks{
		%\IEEEcompsocthanksitem This work was supported by XXX (space reserved).
		\IEEEcompsocthanksitem M. H. Meng is with the
		% School of Computing, 
		National University of Singapore, and also with the Institute for Infocomm Research, A*STAR, Singapore.\protect\\
		E-mail: menghs@i2r.a-star.edu.sg 
		\IEEEcompsocthanksitem G. Bai is with 
		% School of Information Technology and Electrical Engineering, 
		The University of Queensland, Australia.%\protect\\ 
		%Email: g.bai@uq.edu.au
		\IEEEcompsocthanksitem S. G. Teo is with the Institute for Infocomm Research, A*STAR, Singapore. %\protect\\ 
		%Email: teo\_sin\_gee@i2r.a-star.edu.sg
		\IEEEcompsocthanksitem Z. Hou is with the
		% School of Information and Communication Technology, 
		Griffith University, Australia. %\protect\\
		%Email: z.hou@griffith.edu.au,  h.bride@griffith.edu.au 
		\IEEEcompsocthanksitem Y. Xiao, Y. Lin and J. S. Dong are with the
		%School of Computing, 
		National University of Singapore.%\protect\\ 
		%Email: dcsliny@nus.edu.sg
		\IEEEcompsocthanksitem G. Bai (Email: g.bai@uq.edu.au) and Y. Xiao (Email: dcsxan@nus.edu.sg) are corresponding authors.%\protect\\ 
		%Email: dcsdjs@nus.edu.sg
	}
	
	 % \thanks{Manuscript received April 19, 2005; revised August 26, 2015.}
 }

% \markboth{Journal of \LaTeX\ Class Files,~Vol.~14, No.~8, August~2015}
\markboth{2022}
{Meng \MakeLowercase{\textit{et al.}}: Adversarial Robustness of Deep Neural Networks - A Survey from a Formal Verification Perspective}
\IEEEtitleabstractindextext{
	\begin{abstract}
		Neural networks have been widely applied in security applications such as spam and phishing detection, intrusion prevention, and malware detection. 
This black-box method, however, often has uncertainty and poor explainability in applications. 
Furthermore, neural networks themselves are often vulnerable to adversarial attacks. 
For those reasons, there is a high demand for trustworthy and rigorous methods to verify the robustness of neural network models. 
Adversarial robustness, which concerns the reliability of a neural network when dealing with maliciously manipulated inputs, is one of the hottest topics in security and machine learning. 
In this work, we survey existing literature in adversarial robustness verification for neural networks and collect \nofwork diversified research works across machine learning, security, and software engineering domains. 
We systematically analyze their approaches, including how robustness is formulated, what verification techniques are used, and the strengths and limitations of each technique. 
We provide a taxonomy from a formal verification perspective for a comprehensive understanding of this topic. 
We classify the existing techniques based on property specification, problem reduction, and reasoning strategies. 
We also demonstrate representative techniques that have been applied in existing studies with a sample model. 
Finally, we discuss open questions for future research.
	\end{abstract}
	
}

\maketitle

% \fi
\IEEEpeerreviewmaketitle

\IEEEraisesectionheading{\section{Introduction}\label{sec:intro}}
\IEEEPARstart{D}{eep} learning is an \emph{artificial intelligence} (AI) technique that is regarded as one of the top technological breakthroughs in computer science~\cite{lecun2015deep}. It imitates the working principle of the human brain in processing data and forming knowledge patterns and produces promising results for various tasks such as image classification~\cite{krizhevsky2012imagenet}, speech recognition~\cite{hinton2012deep}, recommendation~\cite{deng2016deep}, and natural language understanding~\cite{sutskever2014sequence}. 
Nowadays deep learning is increasingly applied in many fields where trustworthiness is critically needed, such as cybersecurity~\cite{vrejoiu2019neural}, autonomous driving~\cite{tian2018deeptest}, and the healthcare industry~\cite{alipanahi2015predicting}. 
Particularly in cybersecurity, deep learning has been used to perform intrusion detection~\cite{Shone2018}, malware detection~\cite{Nix2017}, phishing detection~\cite{Shima2018}, network traffic analysis~\cite{Okonkwo2022}, etc.
Neural networks, or more specifically deep neural networks (DNNs), act as the key technology for deep learning.
Most neural networks are designed to be deployed in practice where an arbitrary input shall be accepted.
This, however, brings a great uncertainty in determining the behavior of a neural network model even if it has been well-trained~\cite{blundell2015weight}. 
As a result, the trustworthiness of neural networks has turned out to be a new challenging topic in the research community, especially for those to be deployed in safety and security-critical systems~\cite{shahrokni2013systematic}. % Removed fairness as we are going to mention it later

The genesis of neural network verification and validation can be traced back to decades ago~\cite{fu2003neural}, in which the verification is referred to as correctness and the validation is referred to as accuracy and efficiency.
Adversarial examples for neural network models were discovered by Szegedy \emph{et al.}~\cite{szegedy2013intriguing} and later Goodfellow \emph{et al.}~\cite{goodfellow2015explaining} systematically introduced an approach to intrude the robustness property of a neural network to make it misclassify a tampered input to a wrong result.
Since then, neural networks are shown vulnerable to adversarial attacks~\cite{stewart2019security}; consequently, research on the security of neural networks experienced explosive growth.
Many research projects study the attack and defense of AI using adversarial training~\cite{jeong2021smoothmix,madry2018towards,leino2021globally,zhang2021towards}. As a result, verifying the robustness of neural network models against adversarial inputs becomes a strong demand in machine learning research to earn trust from their users and investors.

Thanks to a complete knowledge system of program analysis that has been built since the last century, many techniques designed for traditional program verification, such as \emph{constraint solving}~\cite{katz2017reluplex}, \emph{reachability analysis}~\cite{ruan2018reachability}, and \emph{abstract interpretation}~\cite{gehr2018ai2}, are adopted in reasoning neural network models. 
On the other hand, researchers also attempt to reduce the verification to an optimization problem and achieve uplifting progress by taking advantage of various mathematics theories~\cite{wong2018provable,raghunathan2018certified,dvijotham2018dual}.
The progress made in robustness verification techniques not only increases industry practitioners' confidence in the security of neural network models but also stimulates the discussion of more diverse property specifications in deep learning, such as \emph{fairness}~\cite{albarghouthi2017fairsquare}, \emph{monotonicity}~\cite{seshia2018formal}, and \emph{coverage}~\cite{sun2018concolic}.

An early survey under this theme~\cite{xiang2018verification} performs a comparative study on various verification methods, which collects seven different approaches from the literature and discusses the advantages as well as the shortcomings of each approach. 
A recent survey~\cite{liu2020algorithms} extensively studies 18 existing robustness verification approaches and reproduces them on a unified mathematical framework, therefore for the first time comprehensively sketching the benchmarking of various verification techniques. 
However, some latest progress since 2019 has not been covered.
Another survey~\cite{li2020sok} proposes a taxonomy of robustness verification in accordance with existing robust training techniques and conducts a unified evaluation toolbox for over 20 approaches. It takes both deterministic and probabilistic verification into account and unveils the landscape of available techniques for robustness training and robustness verification.
Compared with existing literature, this paper covers the state-of-the-art robustness verification techniques that have been published in recent years. Moreover, we also propose a more comprehensive taxonomy to categorize the existing approaches and analyze them from a formal verification perspective, thereby forming our insight into future research on this topic.

\subsubsection*{\textbf{Contributions}}
In summary, our paper makes the following contributions:

\begin{itemize}[leftmargin=1.3em]
	\item This survey discusses adversarial robustness verification for deep neural networks from a perspective of formal verification including three key components: a property formalization, a reduction framework, and available reasoning strategies.
	From this point of view, we show the verification methodology not only differs in reasoning strategies but also varies with the author’s understanding of neural network verification as a problem.
	\item We propose a taxonomy based on different components under the formal verification framework and accordingly perform a classification of the existing literature. 
	We establish the connection between different types of approaches from the technical perspective and thereby unveil the landscape as well as the latest progress.
	We particularly analyze the strategy behind the implementation of diversified approaches and discuss their advantages and shortcomings. 
	\item Considering many existing approaches are proposed for specific use cases or experimental settings, it is hard to directly analyze them in a comparative context.
	This survey fills the gap by assessing them on a more general neural network model with a case study supported by substantial visualization.
	\item We suggest the future directions of deep neural network verification, including, improvement in completeness of verification, adoption of various types of complex models, leveraging optimization algorithms, and a generic verification framework.
\end{itemize}

\subsubsection*{\textbf{Scope of This Survey}} 
% Property verification of deep neural networks has been extensively studied during the past decade. 
% For the purpose of this survey, we aim to gather as many relevant works as possible. 
Considering the research on deep neural network property verification begins with the topic of adversarial robustness, which is indeed the most discussed property so far, we choose robustness verification as the theme to carry out our survey.
In order to avoid information overload and present this survey in an appropriate space, we only focus on reviewing and analyzing research works that put \emph{robustness verification} as the major contribution. 

We collect the existing literature published in machine learning, security, and software engineering domains in recent years. We also resort to two sources to replenish the literature. First, we take reference from the existing survey papers on neural network verification~\cite{xiang2018verification,li2020sok,liu2020algorithms}, and thereby we have collected some representative approaches. so that the representative approaches are not missed. Second, we take them as the seeds and treat the counterparts that have been cited and/or benchmarked in their papers as the second source.
% We note that we have only collected those that have been published; therefore, preprints that have not undergone peer review are excluded.

We also notice that most related studies simplify the context of a deep neural network as a \emph{multilayered perceptron} (MLP) for classification tasks that are equipped with three typical activations (ReLU, sigmoid, and tanh), therefore we only discuss robustness verification of MLPs with three activations applicable in this survey. 
To avoid ambiguity, the term ``neural networks'' is referred to as deep neural networks with the assumption that it is in a multi-layer architecture rather than a single neuron perceptron. 
We also assume the robustness verification is performed in a \emph{white-box} setting, which means the architecture and parameters of the examined model are transparent to the verifier.

As a result, we collect \nofwork research works that have proposed an algorithm or a tool to verify the robustness property, and we show them later in Table~\ref{tab:table-all}.
For some research works that do not discuss robustness verification based on a predefined specification, but from the perspective of robust training or finding adversarial examples, we only take their methodologies that apply to specification-based verification into account.

\begin{comment}
\subsubsection*{\textbf{Roadmap}} 
The rest of this paper is organized as follows:

\begin{itemize}
	\item We present relevant background in neural networks and formal verification, followed by our problem definition in Section~\ref{sec:background}.
	We then introduce a formal definition of robustness property in Section~\ref{sec:properties}.
	\item In Section~\ref{sec:correctness}, we introduce the correctness assessment and verifier's capability for neural network property verification. Meanwhile, we discuss the main challenge that past research works are commonly faced with, which is the trade-off between completeness and scalability.
	\item Next, we present our survey result that covers \nofwork research works. In Section~\ref{sec:reduction} we introduce our classification on the reduction of neural network models for the property verification purpose, and right after that we discuss the corresponding reasoning strategies in Section~\ref{sec:reasoning}. 
	\item In the end, we share our takeaways and potential work for the future in Section~\ref{sec:discussion} and conclude this paper in Section~\ref{sec:conclusion}. 
\end{itemize}
\end{comment}

\section{Background}
\label{sec:background}
This section introduces formal verification and one of its typical implementations called \emph{model checking}, including its key elements and processes. 
After that, we propose our definition of property verification by fitting the setting of neural networks into the framework of formal verification. 

\subsection{Formal Verification}
\label{sec:model-checking}

\begin{figure}[t!]
	\centering
	% Trim is performed in the order of {<left> <lower> <right> <upper>}
	\includegraphics[trim=0.1cm 16cm 0cm 0cm,clip=true,width=1\linewidth]{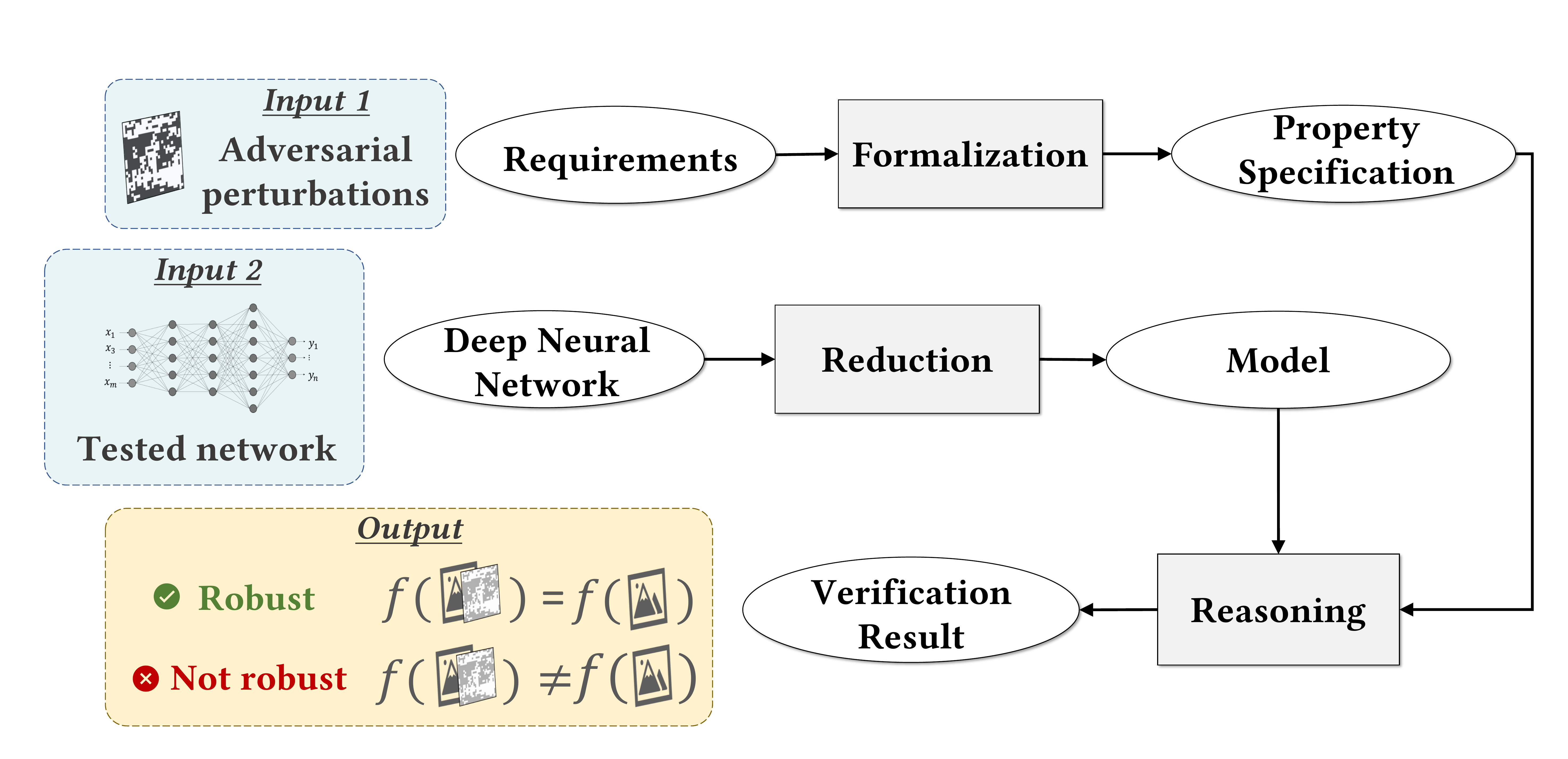}
	\vspace{-15pt}
	\caption{An overview of adversarial robustness verification from a formal verification perspective}
	\label{fig:model-checking}
\end{figure}

Formal verification is defined as a method to prove or disprove the correctness of a specific software or hardware system through formal methods of mathematics. %~\cite{alok2010what}.
As traditional testing is commonly used for finding bugs through an actual execution or simulation, formal verification emphasizes proving the correctness in a modeling framework referring to the examined system with regard to a certain specification or property -- and therefore formal verification requests an in-depth understanding of the system to be examined, and usually comes with a higher cost than traditional testing~\cite{hillel2019why}.
Model checking is a typical formal verification technique to systematically verify if a formally defined property holds in a finite state model~\cite{baier2008principles}. It is designed to explore all possible system states exhaustively and, in the end, provides a qualitative result for specifying whether the target specification holds in the system.
In Fig.~\ref{fig:model-checking} we present a classical schematic view of formal verification through model checking, which is composed of three key components: a \emph{property formalization}, a \emph{reduction framework}, and \emph{specification reasoning}. 

A precise and unambiguous property specification is necessary for formal verification. 
Property specification is a statement of property obtained through formalizing the system requirements. It is often presented as a \emph{classical logic} formula to specify the relationship or restriction on different system components or illustrated in a \emph{temporal logic} formula to enforce a rule over the system in terms of time.
The reduction framework is a formal representation of the system to be examined, which keeps all the key features and functions to ensure a proper simulation of all possible states within the system. 
To sum up, the property formalization defines \emph{what} the system is expected to do, and the system modeling after reduction prescribes \emph{how} the system behaves. 
Property formalization and a proper reduction framework together constitute the \emph{specification phase} of the verification.
These two processes illustrate our understanding of system verification as a \emph{problem}. Different understandings of a system may result in different reasoning approaches available for us to solve the problem. 

With both system and property formalized during the specification phase, the verifier performs the reasoning task and outputs a result -- we call this stage the \emph{verification phase}.
A positive output means that the property specification is satisfied in the examined model. However, a negative output sometimes could not be straightforwardly interpreted as a violation of that property specification, and therefore an extra analysis is required. 
Since the reasoning usually produces a counterexample together with a negative output, we need to determine if the counterexample generated through the simulation faithfully reflects the situation of the actual system -- otherwise, it is concluded as a \emph{false negative} and another round of reasoning is needed after \emph{refining} the reduction framework. Such a strategy to recursively refine the system specification according to the counterexample generated from the previous round is formally proposed in~\cite{clarke2000counterexample} as the counterexample-guided abstraction refinement (CEGAR) approach.
  
\subsection{Property Verification of Deep Neural Networks}

After introducing the basic concept of neural networks and formal verification theory, we can define the problem of property verification of deep neural networks through the lens of formal verification. 
Here we treat a deep neural network model as a function that maps an input to an output, thereby we formalize the property specification as a logic formula involving both input and output.
Given the property specification and available modeling framework on hand, we can perform the verification using diverse reasoning strategies.

Let us take the adversarial robustness as an example. Suppose we have a neural network $f:X\rightarrow Y$ that takes an input $x$ and then outputs $y$ as the result of classification. With a limit of the maximum input perturbation $\Delta_{x}$ given, we assert the neural network is not robust throughout the verification if we observe an arbitrary $(x,y)$ from the training set such that $f(x+\Delta_{x})\ne y$; otherwise the robustness is proved to be held in the given neural network. 

Through the verification above, we can know if a property holds in the target model from the perspective of \emph{the worst case}. However, we gain no knowledge of the percentage of compromised inputs (that are successfully manipulated by the attacker) within the entire input space, not even to mention the analysis of possible approaches to improve. 
For those reasons, \emph{quantitative reasoning} is required to perform a more complex and in-depth property verification for neural network models.
Counting the desirable outputs and offering a probabilistic result is a typical step of quantitative verification that makes it differ from qualitative verification. 
For those properties with demographic characteristics, such as fairness, quantitative reasoning could achieve better semantic expressiveness than any qualitative approach. 
Since we focus on robustness verification, we assume the property specification is qualitatively defined in this survey.

\section{Property Formalization}
\label{sec:properties}
The properties of a neural network could be captured from the semantic context of its specification. 
Most of the properties are \emph{input-output (IO) properties} that request a specific mapping relation between the input and output. 
\emph{Robustness} is one of the earliest IO properties that has been studied, which requests the model's output to be stable upon minor modifications have been made to the input's value~\cite{carlini2017towards,hendrycks2019benchmarking}. 
In this section, we present our definition of adversarial perturbations and provide a formalization of the robustness property.

\subsection{An Introduction to Perturbation}

The perturbation of input can be categorized into two types, namely \emph{global perturbation} and \emph{regional perturbation}, as illustrated in Fig.~\ref{fig:perturbation} (middle). As its name implies, the global perturbation is a vector of the same size as input samples, and the adversarial input is obtained by overlaying the perturbation on a benign input. The source of global perturbation can be from normal distortion in actual deployments such as signal noise or weather conditions~\cite{xiao2021self}, or the output of attack methods, such as \emph{projected gradient descent} (PGD)~\cite{madry2018towards} or \emph{fast gradient sign method} (FGSM)~\cite{goodfellow2015explaining}. 
Regional perturbation is usually a fixed pattern to be added to the original input, which is also known as local perturbation or ``Trojan trigger'' in an attack scenario.
Due to the uncertainty of the occurrence of the attack, verification of security against Trojan attacks will be difficult without knowing the pattern, location, and size of triggers.
Moreover, proving robustness against a specific Trojan pattern does not generalize the property against another Trojan pattern even when both patterns are on a similar scale. For those reasons, regional perturbation is mainly studied in attack scenarios. We focus on global perturbation in the remainder of this paper.

\subsection{Perturbation Measurement}

There are different ways to measure the magnitude (scope) of input perturbations. In mathematics, a \emph{norm} is a function to map a vector's distance to the origin into a non-negative value. Here we treat input perturbations as vectors, and we adopt the norm function to define the scope of perturbation, and correspondingly calculate its magnitude. 

\begin{figure}[t!]
	\centering
	% Trim is performed in the order of {<left> <lower> <right> <upper>}
	\includegraphics[trim=0.1cm 9.5cm 0cm 0.2cm,clip=true,width=1\linewidth]{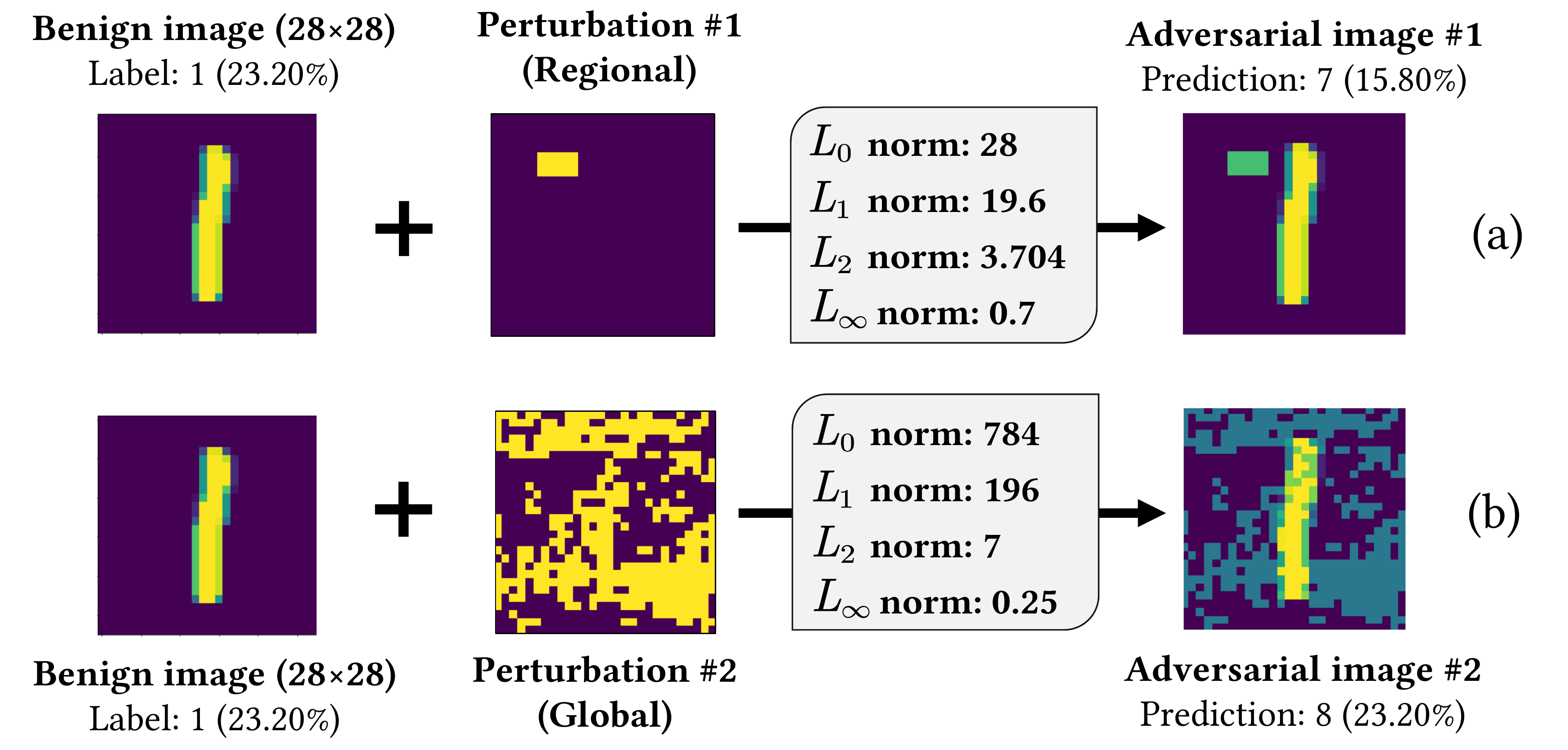}
	\vspace{-17pt}
	\caption{Samples of regional (\textbf{a}) and global (\textbf{b}) adversarial perturbations and measurement of magnitude, illustrated with attacks to the prediction of the MNIST dataset}
	\vspace{-5pt}
	\label{fig:perturbation}
\end{figure}

Although both global and regional perturbations are created with the same objective, which is to maximize the chance of misclassification meanwhile ensuring they are visually indistinguishable, the measuring methods for different types of perturbations may differ. $L_{0}$ norm calculates the number of input features that have been affected, i.e., the size of perturbation, which is useful to describe the size of a regional perturbation. $L_{1}$ norm (Manhattan distance) and $L_{2}$ norm (Euclidean distance) measure the distance between the benign and adversarial inputs, which are often used to describe the accumulated influence of the perturbation. $L_{\infty}$ norm of a perturbation records the greatest magnitude among all elements of it as a vector. $L_{\infty}$ norm is widely used in robustness verification because it echoes the $\epsilon$ value of the FGSM~\cite{goodfellow2015explaining}, which regulates the scope of gradient as the worst-case perturbation that leads to misclassification. Fig.~\ref{fig:perturbation} demonstrates a sample perturbation in each type given a benign input sample from the MNIST dataset~\cite{lecun2010mnist} and illustrates the value of different norm functions in defining the magnitude of input perturbation.

We highlight that the adversarial examples are not the only source of input perturbations, which may also include benign corruption (e.g., the influence of weather) and reasonable distortion (e.g., signal noise, blur)~\cite{hendrycks2019benchmarking}.
However, those cases are specially introduced in object recognition applications and do not differ from the adversarial example in the context of formal analysis.
We assume adversarial examples as the default violation of robustness in this work.

\subsection{Robustness against Adversarial Perturbations}
\label{subsec-robustness}
 
Robustness describes the \emph{fault-tolerant capability} of a computer system to deal with erroneous inputs during execution.
In machine learning, robustness is usually described as the capacity of a model to consistently produce desirable output given some perturbations (or distortions) have been made on the input.
As robustness is closely related to the reliability and performance of a neural network model, it is widely discussed in diverse branches of deep learning applications\cite{hendrycks2019benchmarking}. 

A mutant input with perturbation that can alter the model's output is called an \emph{adversarial example} -- therefore the evaluation of the robustness of a deep neural network model is to find the existence of such an adversarial example.
In analysis, robustness only makes sense if the scope of perturbation is clearly defined and the benign input is properly specified.
The verification may not be able to truthfully reflect the real robustness of the given neural network model if the scope of perturbation is defined negligibly (i.e., too small).
A too strong perturbation makes the adversarial example sufficiently differs from the original input, in which case it is reasonable for the model to classify the input as a different output. 
As for the benign input, it could be an image used for classification or a vector of values within a continuous domain.
We then formalize the robustness property of neural network models as follows:

\begin{definition}[Robustness against adversarial perturbations] \label{def1}
	Given a neural network model that maps a benign input $x_a$ to the output $y_a$, and moreover produces an output $y_b$ given an adversarial input $x_b$. 
	Let $\lVert \cdot \rVert{_p}$ be the function to calculate the $L_p$-norm distance.
	The robustness property $\Phi$ against any adversarial perturbation within the scope of $\Delta_x$ is defined as follows:
	$$
	\Phi(x_a,y_a,x_b,y_b,\Delta_x) \overset{\mathrm{def}}{=} \left(\lVert x_a - x_b \rVert_{p} \le \Delta_x \right) \rightarrow \left(y_a=y_b\right)
	$$
\end{definition}

Other than the above definition that applies to the classification of a discrete domain, the robustness property could be further defined with \emph{tolerant misclassification} or \emph{tolerant misprediction}~\cite{seshia2018formal}, which is more practical in the case that the output falls in a continuous domain (e.g., regression tasks) or an aggregation of labels (e.g., top-K classification). 
As this paper focuses on the classification scenario, we assume the verification does not tolerate errors in the output.

% The 2-page table containing survey result is placed in the section below for the typesetting purpose
\section{Correctness Assessment}
\label{sec:correctness}
From the view of program analysis, the objective of a static checker is to find if a program prevents something that it claims to avoid. The same idea also applies to the verification of neural network models. 
Intuitively, a positive verification outcome implies the neural network model could always prevent violations of specific property if it claims to do so and, on the contrary, a negative outcome claims the neural network model fails to preserve the robustness property.  
However, inconsistency between the claim and the fact sometimes happens when the verification is not guaranteed to be sound and complete. This section discusses \emph{soundness} and \emph{completeness} for a more comprehensive and precise correctness assessment~\cite{michael2017what}.

\subsection{Soundness and Completeness of the Verification}

A proof method is sound if every statement it can prove is indeed true. It is complete if every true statement can be proved.
Given a neural network verifier with something undesired $\mathcal{X}$ that we wish to prevent. The verifier is sound if it never accepts $\mathcal{X}$ taking place in the neural network with certain inputs. It is complete if it never rejects any input, provided $\mathcal{X}$ does not happen in the neural network model.

\begin{figure}[t!]
	\centering
	% Trim is performed in the order of {<left> <lower> <right> <upper>}
	\includegraphics[trim=0cm 12.5cm 0cm 0.2cm,clip=true,width=0.98\linewidth]{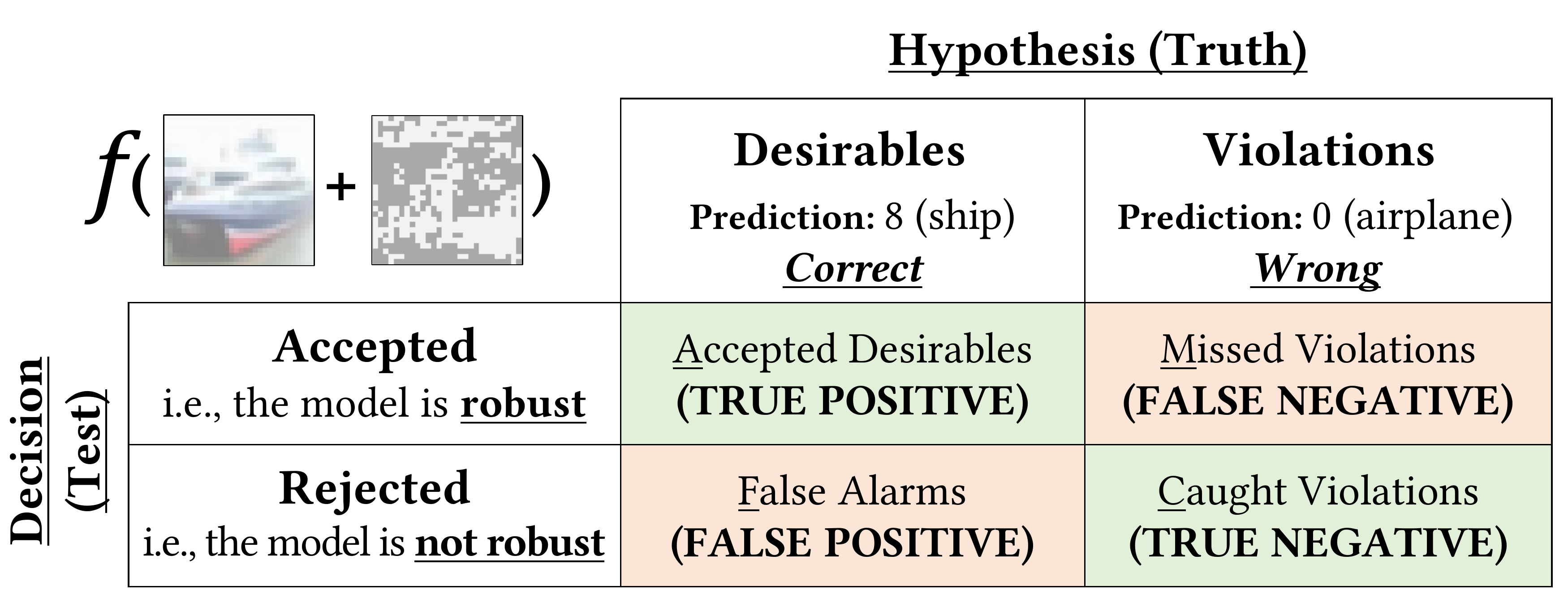}
	\vspace{-5pt}
	\caption{The cases of verification outcomes classified in the statistical hypothesis testing approach, illustrated with the classification of the CIFAR-10 dataset}
	\vspace{-5pt}
	\label{fig:error-chart}
\end{figure}

Statistical hypothesis testing provides another good way to understand soundness and completeness, where the \emph{hypothesis} represents the actual existence of an undesirable event $\mathcal{X}$ defined in the specification of a neural network model, and the \emph{decision} indicates the actual outcome of the verifier~\cite{bertrand2019soundness}. A \emph{table of error} specified for the verification is shown in Fig.\ref{fig:error-chart} with positive and negative hypothesis columns being replaced by desirables and violations.
With such a framework being set up, we may understand the relationship between soundness and completeness as follows:

\begin{itemize}[leftmargin=1.25em]
\item A \textbf{sound} verification prevents \emph{missed violations} (false negatives) but may tolerate \emph{false alarms} (false positives) in case the event $\mathcal{X}$ does not actually take place even if the verifier outputs a rejection. Simply put, the verification may report \emph{unsafe} regarding the (possible) existence of violation $\mathcal{X}$, even though the system is indeed safe. 

\item A \textbf{complete} verification prevents false alarms (false positives). However, it may not be able to account for missed violations (false negatives) — which means there may exist some violations not found by the verification. The verifier may report the system is safe due to its failure to find the violation $\mathcal{X}$.

\item A \textbf{sound} and \textbf{complete} verification implies a \emph{perfect prediction}, that only allows \emph{accepted desirables} (true positives) and \emph{caught violations} (true negatives).
\end{itemize}

\subsection{Trade-off in Practice}

Verifying deep neural network models is shown as an \emph{NP-hard} problem in~\cite{katz2017reluplex} and therefore it is challenging to implement verification that is simultaneously sound, complete, and able to terminate in a reasonable time. 

In practice, one of those three requirements is usually put aside to guarantee the other two characters. 
Some approaches (e.g.,~\cite{katz2017reluplex,ehlers2017formal}) put aside termination to ensure both soundness and completeness are preserved, which is widely adopted in the early stage of neural network verification due to the size of neural networks is comparably small, and the scale of verification is limited. However, scalability becomes a critical factor in the verification of deeper and more complicated neural networks. 
Considering that soundness is more essential in program analysis (to prevent what the system claims to prevent), recent verification solutions (e.g.,~\cite{singh2019abstract,yisrael2019abstraction}) commonly choose to sacrifice completeness to secure soundness and scalability. 

Unlike the complete verification that comes with an exponential time complexity for the worst case, incomplete verifiers usually come with much better scalability but risk the verification from precision loss due to over-approximation. Since the precision loss is accumulated layer by layer, in the worst case, an incomplete verifier may fail to certify the robustness even though it terminates quickly. 

\subsection{Soundness of Floating-point Arithmetic}

The soundness of floating-point arithmetic is another factor to determine the correctness of the verification. Floating-point is a commonly used data type in computing that offers flexible demands in precision. It does not naturally satisfy the axioms of real arithmetic, such as associativity and distributivity~\cite{singh2018fast}. For that reason, the floating-point soundness is a concern in verifying neural network models on a larger scale. Unsound floating-point arithmetic can make the verification suffer from false negatives. At the moment of writing this survey, only a few papers~\cite{singh2018fast,wang2018formal,singh2019abstract} take floating-point arithmetic into consideration. Since we focus on studying robustness verification methodology from a formal verification perspective, floating-point arithmetic is not included in our classification criteria.

% THE BIG TABLE IS PRESENTED BELOW
\begin{table*}[t]
	\centering
	\caption{\label{tab:table-all} 
	A list of the surveyed works in neural network robustness verification.}
	\vspace{-8pt}
	% Config the row spacing
	\def\arraystretch{1.25}
	% Config the column spacing
	\setlength{\tabcolsep}{1.25pt}
	\setcounter{rownumber}{0}
	{\fontsize{6}{7} \selectfont 
	\begin{tabular}[l]{l|l|l|l|c|c|c|c|c|c|c|c|c|c|c|c|c|c|c}			
			\hline 
			\multirow{3}{*}{\textbf{\#}} & \multicolumn{6}{l|}{\textbf{Summary}}                                                                                                                                                                                                                    & \multicolumn{11}{l}{\textbf{Classification\textsuperscript{$\dagger$}}} \\ \cline{2-19} 
			& \multirow{2}{*}{\textbf{\makecell[l]{Authorship, \\Year \& Reference}}} & \multirow{2}{*}{\textbf{\makecell[l]{Name of the Tool \\ \& Link (if any)}}} & \multirow{2}{*}{\textbf{\makecell[l]{Supported\\Framework}}} & \multicolumn{3}{l|}{\textbf{\makecell[l]{Supported\\Activation}}} & \textbf{\makecell[l]{Property}} & \multicolumn{6}{l|}{\textbf{Reduction}} & \multicolumn{5}{l}{\textbf{\makecell[l]{Reasoning\\Strategies}}} \\ \cline{5-19} 
			& \multicolumn{1}{l|}{} & \multicolumn{1}{l|}{} & \multicolumn{1}{l|}{} & \textbf{ReLU} & \textbf{Sigmoid} & \textbf{Tanh} & \multicolumn{1}{c|}{\textbf{Specification}}  & \textbf{SAT} & \textbf{MILP}  & \textbf{SDP} & \textbf{LWP} & \textbf{DI} & \textbf{AI} & \textbf{CR} & \textbf{MO} & \textbf{IA} & \textbf{LA} & \textbf{IR} \\ \hline

% For each row, the first 3 (circ) are for supported activations, after the specs the next 5 are for reductions, and the remaining 5 are for reasoning.

\refstepcounter{rownumber}\therownumber & Pulina \& Tacchella, 2010~\cite{pulina2010abstraction} & \textsc{neVer}					
	& Shark (C++)
	& \emptycirc	& \fullcirc	& \emptycirc	& $L_{\infty}$ 
	& \fullcirc & \emptycirc & \emptycirc  & \emptycirc & \emptycirc & \emptycirc 
	& \fullcirc & \emptycirc & \emptycirc & \fullcirc & \fullcirc\\ \hline	

\refstepcounter{rownumber}\therownumber & Scheibler \textit{et al.}, 2015~\cite{scheibler2015towards} & -- 					
	& Not Specified
	& \emptycirc	& \fullcirc & \emptycirc & IP 
	& \emptycirc & \emptycirc & \emptycirc  & \fullcirc & \emptycirc & \emptycirc 
	& \fullcirc & \emptycirc & \emptycirc & \emptycirc & \emptycirc\\ \hline

\refstepcounter{rownumber}\therownumber & Bastani \textit{et al.}, 2016	\cite{bastani2016measuring}	& \unofficialname{ILP}
	& Caffe
	& \fullcirc & \emptycirc & \emptycirc	& $L_{\infty}$ 
	& \fullcirc & \emptycirc & \emptycirc  & \emptycirc & \emptycirc & \emptycirc 
	& \fullcirc & \emptycirc & \emptycirc & \emptycirc & \emptycirc\\ \hline

\refstepcounter{rownumber}\therownumber & Cheng \textit{et al.}, 2017~\cite{cheng2017maximum}	& \unofficialname{MaxResilience}					
	& ConvNetJS
	& \fullcirc & \emptycirc & \emptycirc	& $L_{1}$ 
	& \emptycirc & \fullcirc & \emptycirc  & \emptycirc & \emptycirc & \emptycirc 
	& \fullcirc & \emptycirc & \fullcirc & \emptycirc & \emptycirc\\ \hline

\refstepcounter{rownumber}\therownumber & Ehler, 2017~\cite{ehlers2017formal} & \textsc{Planet}~\textsuperscript{}	
	& Caffe %\footnote{\url{https://github.com/progirep/planet}}
	& \fullcirc & \emptycirc	& \emptycirc & IP 
	& \fullcirc & \emptycirc & \emptycirc  & \emptycirc & \emptycirc & \emptycirc 
	& \fullcirc & \emptycirc & \fullcirc & \fullcirc & \emptycirc\\ \hline

\refstepcounter{rownumber}\therownumber & Huang \textit{et al.}, 2017~\cite{huang2017safety} & \textsc{DLV}\textsuperscript{}	
	& Keras \,(Theano)%~\footnote{\url{https://github.com/verideep/dlv}}
	& \fullcirc	& \fullcirc  &  \emptycirc	& $L_{1}$, $L_{2}$
	& \emptycirc & \emptycirc & \emptycirc  & \emptycirc & \fullcirc & \emptycirc  
	& \fullcirc & \emptycirc & \emptycirc & \emptycirc & \emptycirc\\ \hline

\refstepcounter{rownumber}\therownumber & Katz \textit{et al.}, 2017	\cite{katz2017reluplex}	& \textsc{Reluplex}\textsuperscript{}	
	& Not Specified (C)%~\footnote{\url{ht tps://github.com/guykatzz/ReluplexCav2017}}
	& \fullcirc & \emptycirc	& \emptycirc & IP 
	& \fullcirc & \emptycirc & \emptycirc  & \emptycirc & \emptycirc & \emptycirc 
	& \fullcirc & \emptycirc & \emptycirc & \emptycirc & \fullcirc\\ \hline

\refstepcounter{rownumber}\therownumber & Lomuscio \& Maganti, 2017~\cite{lomuscio2017approach}& \unofficialname{NSVerify}
	& Not Specified
	& \fullcirc & \emptycirc	& \emptycirc 	& IP 
	& \emptycirc & \fullcirc & \emptycirc  & \emptycirc & \emptycirc & \emptycirc 
	& \fullcirc & \emptycirc & \emptycirc & \emptycirc & \emptycirc\\ \hline

\refstepcounter{rownumber}\therownumber & Wong \& Kolter, 2018~\cite{wong2018provable} & \makecell[lt]{\textsc{ConvexAdversarial}, \unofficialname{ConvDual}\textsuperscript{}	}%,\\ \unofficialname{LP-Full}}		
	& PyTorch %\footnote{\href{https://github.com/locuslab/convex_adversarial}{https://github.com/locuslab/convex\_adversarial}}
	& \fullcirc & \emptycirc	& \emptycirc & $L_{\infty}$
	& \fullcirc & \emptycirc & \emptycirc  & \emptycirc & \emptycirc & \emptycirc 
	& \emptycirc & \fullcirc & \emptycirc & \fullcirc & \emptycirc\\ \hline

\refstepcounter{rownumber}\therownumber & Xiang \textit{et al.}, 2017~\cite{xiang2017reachable} & \unofficialname{ExactReach}	
	& MATLAB
	&\fullcirc & \emptycirc & \emptycirc & IP 
	& \emptycirc & \emptycirc & \emptycirc  & \fullcirc & \emptycirc & \emptycirc  
	& \emptycirc & \emptycirc & \fullcirc & \emptycirc & \emptycirc \\ \hline

\refstepcounter{rownumber}\therownumber & Dutta \textit{et al.}, 2018~\cite{dutta2018output}	& \textsc{Sherlock}\textsuperscript{}			
	& Not Specified (C++)%~\footnote{\url{https://github.com/verivital/sherlock}}
	 & \fullcirc & \emptycirc	& \emptycirc 	& $L_{\infty}$ 
	 & \emptycirc & \fullcirc & \emptycirc  & \emptycirc & \emptycirc & \emptycirc 
	 & \fullcirc & \emptycirc & \emptycirc & \emptycirc & \emptycirc\\ \hline

\refstepcounter{rownumber}\therownumber & Dvijotham \textit{et al.}, 2018~\cite{dvijotham2018dual}	&  \unofficialname{DeepVerify}, \unofficialname{Duality}\textsuperscript{}						
	& TensorFlow %~\footnote{\url{https://github.com/deepmind/deep-verify}}
	& \fullcirc	& \fullcirc		& \fullcirc			 & $L_{\infty}$
	& \fullcirc & \emptycirc & \emptycirc  & \emptycirc & \emptycirc & \emptycirc 
	& \emptycirc & \fullcirc & \emptycirc & \fullcirc & \emptycirc\\ \hline

\refstepcounter{rownumber}\therownumber & Raghunathan \textit{et al.}, 2018~\cite{raghunathan2018certified}	&  \unofficialname{AdvDefense}, \unofficialname{Certify}\textsuperscript{}						
	& TensorFlow %\footnote{\url{https://worksheets.codalab.org/worksheets/0xa21e794020bb474d8804ec7bc0543f52/}}
	& \fullcirc	& \fullcirc		& \fullcirc	 & $L_{\infty}$
	& \emptycirc & \emptycirc & \fullcirc & \emptycirc & \emptycirc & \emptycirc 
	& \emptycirc & \fullcirc & \emptycirc & \fullcirc & \emptycirc\\ \hline

\refstepcounter{rownumber}\therownumber & Raghunathan \textit{et al.}, 2018~\cite{raghunathan2018semidefinite}	&  \unofficialname{SDPVerify}\textsuperscript{}						
	& TensorFlow \& MATLAB %\footnote{\url{https://worksheets.codalab.org/worksheets/0x6933b8cdbbfd424584062cdf40865f30/}}
	& \fullcirc	& \emptycirc		& \emptycirc	 & $L_{\infty}$
	& \emptycirc & \emptycirc & \fullcirc & \emptycirc & \emptycirc & \emptycirc 
	& \emptycirc & \fullcirc & \emptycirc & \emptycirc & \emptycirc\\ \hline

\refstepcounter{rownumber}\therownumber & Gehr \textit{et al.}, 2018~\cite{gehr2018ai2}	&  \textsc{AI2}\textsuperscript{}				
	& PyTorch \& TensorFlow %\footnote{\url{https://github.com/eth-sri/eran}}
	& \fullcirc	& \emptycirc	& \emptycirc & $L_{\infty}$
	& \emptycirc & \emptycirc & \emptycirc & \emptycirc & \emptycirc & \fullcirc  
	& \emptycirc & \emptycirc & \fullcirc & \emptycirc & \emptycirc\\ \hline

\refstepcounter{rownumber}\therownumber & Singh \textit{et al.}, 2018~\cite{singh2018fast} 	&  \textsc{DeepZ}\textsuperscript{}			
	& Not Specified (C, Python)
	& \fullcirc	& \fullcirc			& \fullcirc		 & $L_{\infty}$ 
	& \emptycirc & \emptycirc & \emptycirc & \emptycirc & \emptycirc & \fullcirc  
	& \emptycirc & \emptycirc & \fullcirc & \fullcirc & \emptycirc\\ \hline

\refstepcounter{rownumber}\therownumber & Wang \textit{et al.}, 2018~\cite{wang2018formal} 		&  \textsc{ReluVal}\textsuperscript{}			
	& Not Specified (C)%~\footnote{\url{https://github.com/tcwangshiqi-columbia/ReluVal}}
	& \fullcirc	& \emptycirc	& \emptycirc 	 & $L_{\infty}$ 
	& \emptycirc & \emptycirc & \emptycirc & \fullcirc & \emptycirc & \emptycirc  
	& \fullcirc & \emptycirc & \fullcirc & \emptycirc & \fullcirc\\ \hline

\refstepcounter{rownumber}\therownumber & Wang \textit{et al.}, 2018	\cite{wang2018efficient} 	&  \textsc{Neurify}\textsuperscript{}	
	& TensorFlow%~\footnote{\url{https://github.com/tcwangshiqi-columbia/Neurify}}
	& \fullcirc	& \emptycirc	& \emptycirc & $L_{1}$, $L_{2}$, $L_{\infty}$
	& \emptycirc & \emptycirc & \emptycirc & \fullcirc & \emptycirc & \emptycirc  
	& \emptycirc & \emptycirc & \fullcirc & \fullcirc & \fullcirc \\ \hline

\refstepcounter{rownumber}\therownumber & Weng \textit{et al.}, 2018	\cite{weng2018towards} 		&  \textsc{Fast-Lin}, \textsc{Fast-Lip}\textsuperscript{}			
	& TensorFlow%~\footnote{\url{https://github.com/huanzhang12/CertifiedReLURobustness}}
	& \fullcirc	& \emptycirc	& \emptycirc & $L_{1}$, $L_{2}$, $L_{\infty}$
	& \emptycirc & \emptycirc & \emptycirc & \fullcirc & \emptycirc & \emptycirc 
	& \fullcirc & \emptycirc & \fullcirc & \fullcirc & \emptycirc \\ \hline

\refstepcounter{rownumber}\therownumber & Xiang \textit{et al.}, 2018~\cite{xiang2018output} 		&  \unofficialname{MaxSensitivity}						
	& MATLAB
	& \fullcirc	& \fullcirc		& \fullcirc		 	& $L_{\infty}$
	& \emptycirc & \emptycirc & \emptycirc & \fullcirc & \emptycirc & \emptycirc 
	& \emptycirc & \emptycirc & \fullcirc & \emptycirc & \emptycirc \\ \hline

\refstepcounter{rownumber}\therownumber & Zhang \textit{et al.}, 2018~\cite{zhang2018efficient} 	&  \textsc{CROWN}\textsuperscript{}	
	& TensorFlow %~\footnote{\url{https://github.com/crown-robustness/crown}}
	& \fullcirc	& \fullcirc			& \fullcirc & $L_{1}$, $L_{2}$, $L_{\infty}$
	& \emptycirc & \emptycirc & \emptycirc & \fullcirc & \emptycirc & \emptycirc 
	& \emptycirc & \emptycirc & \fullcirc & \fullcirc & \emptycirc \\ \hline

\refstepcounter{rownumber}\therownumber & Bunel \textit{et al.}, 2018~\cite{bunel2018unified}		&  \textsc{BaB, BaBSB \& reluBaB}\textsuperscript{}	
	& PyTorch %\footnote{\url{https://github.com/oval-group/PLNN-verification}}
	& \fullcirc & \emptycirc & \emptycirc & $L_{\infty}$ 
	& \emptycirc & \fullcirc & \emptycirc & \emptycirc & \emptycirc & \emptycirc 
	& \fullcirc & \emptycirc & \fullcirc & \fullcirc & \fullcirc \\ \hline

\refstepcounter{rownumber}\therownumber & Katz \textit{et al.}, 2019	\cite{katz2019marabou}		&  \textsc{Marabou}\textsuperscript{}			
	& TensorFlow%~\footnote{\url{https://github.com/NeuralNetworkVerification/Marabou}}
	& \fullcirc & \emptycirc	& \emptycirc 	& IP
	& \fullcirc & \emptycirc & \emptycirc & \emptycirc & \emptycirc & \emptycirc 
	& \fullcirc & \emptycirc & \emptycirc & \emptycirc & \fullcirc \\ \hline

\refstepcounter{rownumber}\therownumber & Tjeng \textit{et al.}, 2019~\cite{tjeng2019evaluating}	&  \textsc{MIPVerify}\textsuperscript{}	
	& Julia %\footnote{\url{https://github.com/vtjeng/MIPVerify.jl}}
	& \fullcirc & \emptycirc	& \emptycirc 	& $L_{1}$, $L_{2}$, $L_{\infty}$ 
	& \emptycirc & \fullcirc & \emptycirc & \emptycirc & \emptycirc & \emptycirc 
	& \fullcirc & \emptycirc & \emptycirc & \emptycirc & \emptycirc \\ \hline

\refstepcounter{rownumber}\therownumber & Singh \textit{et al.}, 2019~\cite{singh2019abstract} 	&  \textsc{DeepPoly}\textsuperscript{}			
	& PyTorch \& TensorFlow 
	& \fullcirc	& \fullcirc & \fullcirc		 & $L_{\infty}$
	& \emptycirc & \emptycirc & \emptycirc& \emptycirc & \emptycirc & \fullcirc 
	& \emptycirc & \emptycirc & \fullcirc & \fullcirc & \fullcirc \\ \hline

\refstepcounter{rownumber}\therownumber & Singh \textit{et al.}, 2019~\cite{singh2019boosting}		&  \textsc{RefineZono}\textsuperscript{}		
	& PyTorch \& TensorFlow 
	& \fullcirc	& \emptycirc	& \emptycirc & $L_{\infty}$
	& \emptycirc & \fullcirc & \emptycirc & \emptycirc & \emptycirc & \fullcirc 
	& \fullcirc & \emptycirc & \fullcirc & \fullcirc & \fullcirc \\ \hline

\refstepcounter{rownumber}\therownumber & Singh \textit{et al.}, 2019~\cite{singh2019beyond}		&  \textsc{kPoly}, \textsc{RefinePoly}\textsuperscript{}			
	& PyTorch \& TensorFlow 
	& \fullcirc	& \emptycirc	& \emptycirc & $L_{\infty}$ 
	& \emptycirc & \emptycirc & \emptycirc & \emptycirc & \emptycirc & \fullcirc 
	& \fullcirc & \emptycirc & \fullcirc & \fullcirc & \fullcirc \\ \hline

\refstepcounter{rownumber}\therownumber & Yisrael \textit{et al.}, 2019~\cite{yisrael2019abstraction} 	&  \unofficialname{CEGAR-Marabou}					
	& Not Specified
	& \fullcirc	& \emptycirc		&  \emptycirc	& $L_{1}$, $L_{\infty}$ 
	& \emptycirc & \emptycirc & \emptycirc & \emptycirc & \emptycirc & \fullcirc  
	& \emptycirc & \emptycirc & \fullcirc & \emptycirc & \fullcirc\\ \hline

\refstepcounter{rownumber}\therownumber & Huang \textit{et al.}, 2019~\cite{huang2019reachnn} 	&  \textsc{ReachNN}\textsuperscript{}				
	& Not Specific (Python) 
	& \fullcirc	& \fullcirc		&  \fullcirc	& IP
	& \emptycirc & \emptycirc & \emptycirc & \fullcirc & \emptycirc & \emptycirc  
	& \emptycirc & \emptycirc & \fullcirc & \fullcirc & \emptycirc\\ \hline

\refstepcounter{rownumber}\therownumber & Xiang \textit{et al.}, 2020~\cite{xiang2020reachable} 	&  \textsc{IGNNV}\textsuperscript{}				
	& MATLAB %\footnote{\url{https://github.com/xiangweiming/ignnv}}
	& \fullcirc	& \fullcirc		&  \fullcirc	& IP
	& \emptycirc & \emptycirc & \emptycirc & \fullcirc & \emptycirc & \emptycirc 
	& \emptycirc & \emptycirc & \fullcirc & \emptycirc & \emptycirc\\ \hline

\refstepcounter{rownumber}\therownumber & Bak \textit{et al.}, 2020~\cite{bak2020improved} 	&  \textsc{NNenum}\textsuperscript{}				
	& Not Specified (Python)
	& \fullcirc	& \emptycirc & \emptycirc	& IP
	& \emptycirc & \emptycirc & \emptycirc  & \emptycirc & \emptycirc & \fullcirc 
	& \fullcirc & \emptycirc & \emptycirc & \fullcirc & \fullcirc \\ \hline

\refstepcounter{rownumber}\therownumber & Tran \textit{et al.}, 2019 \& 2020~\cite{tran2019star,tran2020nnv} 	&  \textsc{NNV}\textsuperscript{}	
	& MATLAB 
	& \fullcirc	& \fullcirc & \fullcirc	& IP
	& \emptycirc & \emptycirc & \emptycirc & \fullcirc  & \emptycirc & \emptycirc
	& \emptycirc & \emptycirc & \fullcirc & \fullcirc & \emptycirc \\ \hline

\refstepcounter{rownumber}\therownumber & Henriksen \& Lomuscio, 2020~\cite{henriksen2020efficient} 	&  \textsc{VeriNet}\textsuperscript{}	
	& PyTorch
	& \fullcirc	& \fullcirc	& \fullcirc	& $L_{1}, L_{\infty}$
	& \fullcirc	& \emptycirc & \emptycirc & \emptycirc & \emptycirc & \emptycirc 
	& \fullcirc & \emptycirc & \fullcirc & \fullcirc & \fullcirc \\ \hline

\refstepcounter{rownumber}\therownumber & Botoeva \textit{et al.}, 2020~\cite{botoeva2020efficient} 	&  \textsc{Venus}\textsuperscript{}	
	& Keras (TensorFlow)
	& \fullcirc	& \emptycirc & \emptycirc & $L_{\infty}$
	& \emptycirc & \fullcirc & \emptycirc & \emptycirc & \emptycirc & \emptycirc 
	& \fullcirc & \emptycirc & \emptycirc & \emptycirc & \fullcirc	\\ \hline
	
\refstepcounter{rownumber}\therownumber & Dathathri \textit{et al.}, 2020~\cite{dathathri2020enabling} 	&  \textsc{SDP-FO}\textsuperscript{}				
	& JAX %\footnote{\url{https://github.com/deepmind/jax_verify}}
	& \fullcirc	& \fullcirc		& \fullcirc	 & $L_{\infty}$
	& \emptycirc & \emptycirc & \fullcirc & \emptycirc & \emptycirc & \emptycirc 
	& \emptycirc & \fullcirc & \emptycirc & \fullcirc & \emptycirc\\ \hline

\refstepcounter{rownumber}\therownumber & Tjandraatmadja \textit{et al.}, 2020~\cite{tjandraatmadja2020convex} 	&  \textsc{FastC2V}, \textsc{Opt2CV}\textsuperscript{}				
	& Not Specified (C++) %\footnote{\url{https://github.com/google-research/tf-opt}}
	& \fullcirc	& \fullcirc		& \fullcirc	 & $L_{\infty}$
	& \fullcirc & \emptycirc & \emptycirc & \emptycirc & \emptycirc & \emptycirc 
	& \emptycirc & \fullcirc & \emptycirc & \emptycirc & \emptycirc\\ \hline

\refstepcounter{rownumber}\therownumber & Fazlyab \textit{et al.}, 2020~\cite{fazlyab2020safety} 	&  \textsc{DeepSDP}\textsuperscript{}	
	& MATLAB 
	& \fullcirc	& \fullcirc & \fullcirc	& $L_{\infty}$
	& \emptycirc & \emptycirc & \fullcirc & \emptycirc & \emptycirc & \emptycirc 
	& \emptycirc & \fullcirc & \emptycirc & \fullcirc & \emptycirc \\ \hline
	
\refstepcounter{rownumber}\therownumber & Batten \textit{et al.}, 2021~\cite{batten2021efficient} 	&  \textsc{LayerSDP}, \textsc{FastSDP}
	& Not Specified 
	& \fullcirc	& \emptycirc & \emptycirc & $L_{\infty}$
	& \emptycirc & \emptycirc & \fullcirc & \emptycirc & \emptycirc & \emptycirc 
	&  \emptycirc & \fullcirc & \emptycirc & \fullcirc & \emptycirc\\ \hline

\refstepcounter{rownumber}\therownumber & Kouvaros \& Lomuscio, 2021~\cite{kouvaros2021towards} 	&  \textsc{Venus2}\textsuperscript{$\mathsection$}
	& Not Specified 
	& \fullcirc	& \emptycirc & \emptycirc & $L_{\infty}$
	& \emptycirc & \fullcirc & \emptycirc & \emptycirc & \emptycirc & \emptycirc 
	& \fullcirc & \emptycirc & \emptycirc & \emptycirc & \fullcirc	\\ \hline

\hline

		\end{tabular}
		% GRAND TABLE ENDS HERE
	}
\begin{flushleft}
	%\textsuperscript{$\dagger$} We discuss the top three commonly used activation functions in this table. 
	%\vspace{2pt}

	\textsuperscript{$\dagger$} Property specifications include fixed input pattern (IP) or $L_{p}$ norms of adversarial perturbations. Reduction approaches are classified into five classes as shown in the table, including SAT/LP encoding (SAT), MILP encoding (MILP), QCQP/SDP encoding (SDP), layerwise propagation (LWP), discretization (DI), and abstract interpretation (AI). Reasoning strategies are classified into five classes as shown in the table, including constraint reasoning (CR), mathematical optimization (MO), interval arithmetic (IA), linear approximation (LA), and iterative refinement (IR).
	\vspace{2pt}
	
	\textsuperscript{$\mathsection$} The link is provided in the literature but not accessible as of May 2022.
	\vspace{2pt}

	\textsuperscript{*} The name of the tool has not been explicitly defined in the literature but is taken from the source code released, or commonly addressed in subsequent research. The name is usually cited from the function name or the paper title.
\end{flushleft}
\vspace{-10pt}
\end{table*}
% THE BIG TABLE ENDS HERE

\section{Model Reduction}
\label{sec:reduction}
\begin{figure}[!t]
	\centering
	% Trim is performed in the order of {<left> <lower> <right> <upper>}
	\includegraphics[trim=0cm 6.3cm 1cm 5.3cm, clip, width=.8\linewidth]{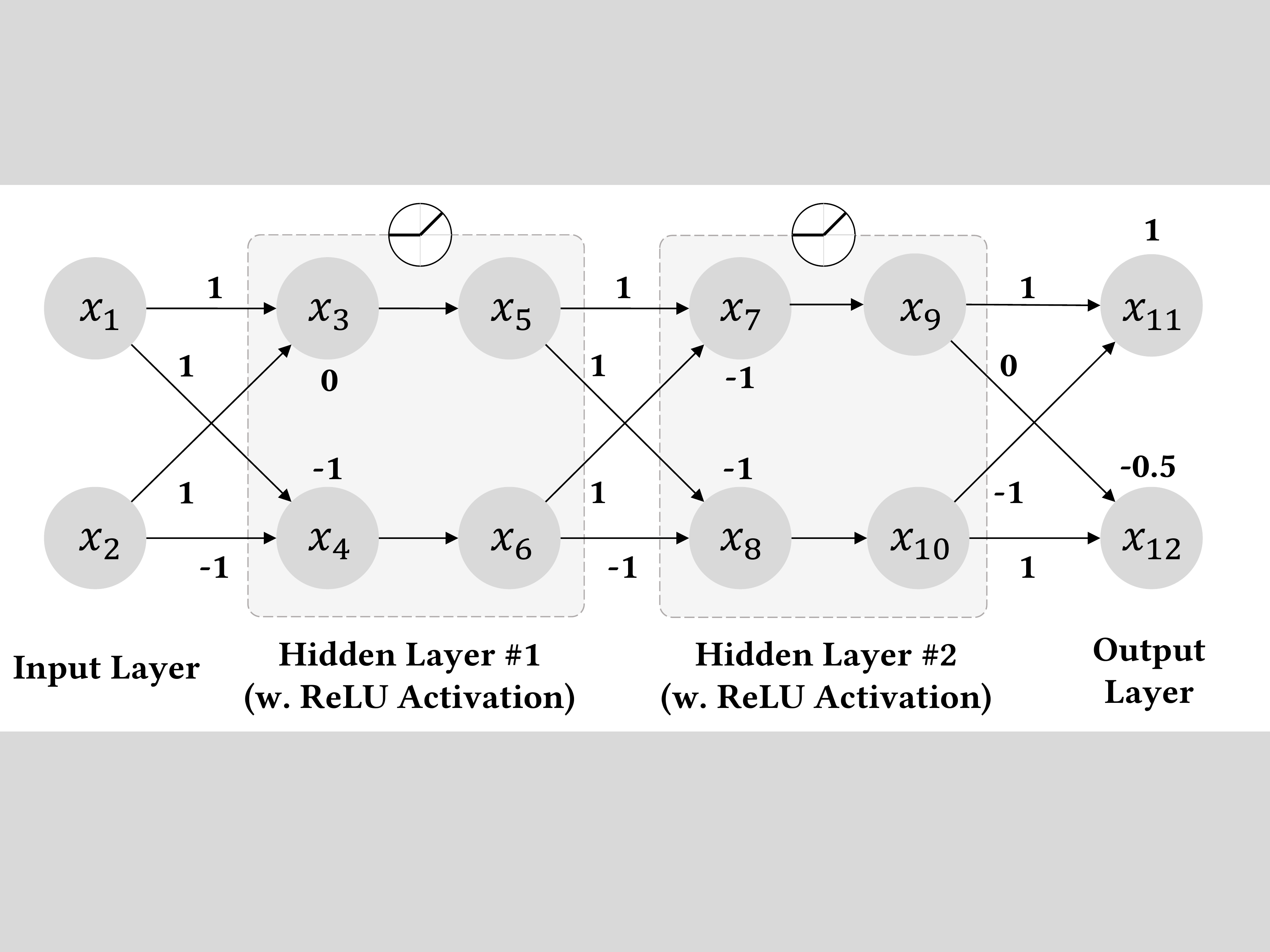}
	\vspace{-5pt}
	\caption{The sample neural network (ReLU activation) with weight and bias parameters shown on edges and nodes respectively}
	\label{fig:sample_network}
\end{figure}

% In previous sections, we have proposed our property formalization and defined the assessment criteria. 
This section discusses how to reduce the property verification of a neural network model to another problem with finite states and an available solution. 
Such a reduction process is also known as \emph{modeling} in model checking.
The verification reduction reflects our understanding of the verification as a problem and meanwhile determines what consequent approach or algorithm could be used to solve it. 
As one of the three key processes defined in Section~\ref{sec:model-checking}, reduction stands for a formal interpretation of the neural network model to be examined. 

We categorize the existing reduction methods into two groups, namely \emph{optimization problem} and \emph{reachability problem}. 
In case the verification is treated as an optimization problem, we further classify the reduction of existing approaches into \emph{SAT/LP encoding}, \emph{MILP encoding}, and \emph{QCQP/SDP encoding}.
For the second category, we also identify three sub-classes from the existing works, namely \emph{layerwise propagation}, \emph{abstract interpretation}, and \emph{discretization}. 
We present our classification result over the collected literature in columns 9-14 of Table~\ref{tab:table-all} and detail those reduction methods in the remainder of this section.

To facilitate the analysis and discussion, we generate a sample neural network model. 
As shown in Fig.\ref{fig:sample_network}, the sample network is a fully-connected four-layer MLP with ReLU activation, which is a typical neural network architecture used for classification tasks. 
To make it clear and consistent, we call the basic elements within the sample neural network as \emph{nodes}. We accordingly separate each neuron within hidden layers into two nodes, which perform an \emph{affine function} and \emph{activation function} respectively.
We create this sample network based on the \emph{fact} that the specific input $[0, 0]$ would always lead to the class corresponding to $x_{11}$.
In addition, we also claim \emph{factual robustness} of the sample network when dealing with $L_{\infty}$ perturbations that are less or equal to $1.0$, i.e., the value of $x_{11}$ is always greater than $x_{12}$ regardless of any input perturbation within the range $[-1, 1]$. 

\subsection{Boolean Satisfiability Problem (SAT) / Linear Programming (LP) Encoding}

Since a neural network model is analyzed in a white-box setting and all the parameters are constants, it is intuitive to perform the analysis by transforming the entire neural network into a set of constraints in the form of \emph{propositional logic} and linking certain constraints containing the input and output in one Boolean formula. 
We aim to prove the non-existence of any violation in the examined model, and thereby we can reduce the robustness verification to a constraint satisfaction problem, or more specifically a Boolean satisfiability problem (colloquially known as SAT problem). 

Alternatively, we can also set the negation of constraints over the output nodes as the \emph{objective function} and keep the remaining constraints unchanged, thus the verification process is transformed into solving an optimization problem. In the case of all activation are \emph{piecewise-linear functions} (e.g., ReLU), this reduction approach is also known as linear programming (LP) encoding. 

As we are checking whether a neural network is robust, the goal of reasoning over this encoding is to prove the existence of such a solution that satisfies all the constraints, rather than finding the objective value. 
To this end, we symbolize every node with a unique variable and then connect those variables layer-by-layer according to the computational relation defined in the network model. This reduction approach (in SAT form) can be formalized as below.

\begin{definition}[SAT/LP encoding] \label{def_reduction_sat}
	Given a neural network model $f$ with $n$ nodes. To verify a property specification of it, $f$ can be formally reduced to a model $\mathcal{M}$ in form of a constraint satisfaction problem, such as:
	$$
	\mathcal{M}_{\text{(SAT/LP)}}=\left<V,D,C\right>   
	$$
	where $V=\left\{V_1,…,V_n \right\}$ is a set of variables corresponding to the nodes within $f$; $D=\left\{D_1,…,D_n \right\}$ is a set specifying the value domain (if any) of variables in $V$; and $C=\left\{C_1,…,C_m \right\}$ is the set of constraints, including the constraint reflecting the violation of property specification.   
	
\end{definition}

Fig.\ref{fig:sat-encoding} depicts the encoding of the sample network that \emph{violates} the robustness specification, i.e., $x_{11} \le x_{12}$. 
Because ReLU is a \emph{piecewise-linear} function that produces different outputs from two input domains, we notice that each ReLU activation function introduces a disjunctive logical (\texttt{or}) operation, which bisects the verification problem. As a result, given $n$ ReLU nodes in a neural network model, its SAT/LP encoding creates a disjunction of $2^{n}$ conjunctive normal form (CNF) formulas (16 in the sample model) in total.

As we capture all nodes from the examined model and truthfully encode both affine and activation functions in constraints, we consider this reduction approach preserves both soundness and completeness for the next step of reasoning.
Nevertheless, the nature of propositional logic is only suitable to encode \emph{linear algebraic relations} without exponential. 
It makes this reduction approach only applicable to neural networks that use piecewise-linear activation, such as ReLU or its variants; otherwise, the output of activation must be relaxed by another linear or piecewise-linear function in advance.

\subsection{Mixed-Integer Linear Programming (MILP) Encoding}

One significant progress made on SAT/LP encoding is achieved by adding an integer variable for each ReLU activation and thereby encodes the examined model to a \emph{mixed-integer linear programming} (MILP) problem. MILP is a variant of LP that allows part of variables specified as integers. Some recent papers~\cite{cheng2017maximum,lomuscio2017approach,dutta2018output,fischetti2019deep,tjeng2019evaluating} take advantage of the integer setting of MILP, propose a subtle encoding of the ReLU function into a set of four in-equations, and thus eliminate the excessive disjunctive formulas in the constraints. 

We show how a canonical integer setting could be adopted in ReLU activation in Definition~\ref{def_reduction_milp_bigm} and present our formalization of MILP encoding in Definition~\ref{def_reduction_milp}. We then demonstrate the corresponding encoding of the sample network in Fig.\ref{fig:milp-encoding}.

\begin{definition}[Big-M encoding of ReLU] \label{def_reduction_milp_bigm}
	Consider given an integer variable $d\in\left\{0,1\right\}$ and another sufficiently great constant $M$ (which is guaranteed to be greater than the upper bound of ReLU input), a ReLU function $y=max\left(0,x\right)$ could be encoded by a CNF of four logical formulas, as shown below:
	\begin{multline*}
	y=max\left(0,x\right) \Leftrightarrow \left( \left( y\ge x \right) \land \left( y\le x+M\cdot d \right)\right. \\
	\left. \land \left( y\ge 0 \right) \land \left( y\le M\cdot \left( 1-d \right) \right)\right)
	\end{multline*}
	where the case $d=0$ represents the ReLU is in \emph{active} mode ($x>0$ and $y=x$), and conversely, $d=1$ indicates the ReLU is \emph{inactive} ($x<0$ and $y=0$).
\end{definition}

\begin{definition}[MILP encoding] \label{def_reduction_milp}
	Given a neural network model $f$ with $n$ nodes. To verify a property of it, $f$ can be formally modeled as a MILP problem presented as follows:
	$$
	\mathcal{M}_{\text{(MILP)}}= \left<V,V_{int},D,C,\text{Obj}\right>   
	$$
	where $V=\left\{V_1,…,V_n \right\}$ is a set of variables corresponding to the nodes within $f$;
	$V_{int}$ is the set of integers that correspondingly defined to constraint ReLU activation outputs; $D=\left\{D_1,…,D_n \right\}$ is a set specifying the value domain (if any) of variables in $V$; $C=\left\{C_1,…,C_m \right\}$ is the set of constraints; and $\text{Obj}$ is the objective (minimum) inequation looking for the counterexample in case of violation (i.e., $\text{Obj} \le 0$) occurs.   
\end{definition}

\begin{figure}[tb!]
	\begin{footnotesize}
		\begin{equation}
			\begin{gathered}
				% \mathcal{M}_{(SAT)}:\left< V,D,C \right>, s.t.\\
				V=\left\{x_1,x_2,...,x_{12}\right\},
				D=\left\{ -1 \le x_1\le 1,\, -1 \le x_2\le 1\right\}
			\end{gathered}
			\nonumber
		\end{equation}	\vspace{-1.3cm}
		\begin{multline*}\\
			C=\left\{ x_3=x_1+x_2,\, x_4=x_1-x_2-1, \right.\\
			\left( \left( x_3\le 0 \right) \land \left( x_5=0 \right) \right) \lor \left( \left( x_3>0 \right) \land \left( x_5=x_3 \right) \right),\\
			\left( \left( x_4\le 0 \right) \land \left( x_6=0 \right) \right) \lor \left( \left( x_4>0 \right) \land \left( x_6=x_4 \right) \right),\\
			\left. x_7=-x_5+x_6-1,\, x_8=x_5-x_6-1, \right.\\
			\left( \left( x_7\le 0 \right) \land \left( x_9=0 \right) \right) \lor \left( \left( x_7>0 \right) \land \left( x_9=x_7 \right) \right), \\
			\left( \left(  x_8\le 0 \right) \land \left( x_{10}=0 \right) \right) \lor \left( \left( x_8>0 \right) \land \left( x_{10}=x_8 \right)\right), \\
			\left. x_{11}=x_9-x_{10}+1,\, x_{12}=x_{10}-0.5,\, x_{11}\le x_{12} \right\}
		\end{multline*}
	\end{footnotesize}
	\vspace{-20pt}
	\caption{Boolean formula obtained through SAT encoding of the sample network given the robustness specification being violated}
	\label{fig:sat-encoding}
\end{figure}

It is worth noting that some existing approaches (e.g.,~\cite{raghunathan2018certified,dathathri2020enabling,batten2021efficient}, etc) initiate their analysis from an LP or MILP perspective and later transform the neural network encoding into a semidefinite programming (SDP) problem. We separate this encoding approach as an independent category and detail it in Section~\ref{sec:sdp-encoding}.

The MILP encoding is proposed as a successor of SAT/LP encoding. It retains both soundness and completeness at the reduction stage. Moreover, it addresses our concern about the ``too many branches'' issue in SAT/LP encoding, especially when handling larger or deeper neural networks, and eventually brings potential improvement regarding the scalability in neural network verification. 

\subsection{Quadratically Constrained Quadratic Program (QCQP) / Semidefinite Programming (SDP) Encoding}
\label{sec:sdp-encoding}

A quadratically constrained quadratic program (QCQP) is an optimization problem in which both the objective function and the constraints are quadratic functions. 
In fact, both LP and MILP encoding can be treated as variants of QCQP where the quadratic objective function and quadratic constraints are absent.
Solving a QCQP is NP-hard, too. However, the problem can be solved by semidefinite programming (SDP) if it is convex. 
For that reason, existing studies categorized in this type are also known as SDP-based approaches~\cite{dathathri2020enabling,liu2020algorithms,raghunathan2018semidefinite}.
They mainly transform the ReLU activation to convex quadratic equations, thereby the verification can be reduced to a convex QCQP after including the linear constraints (i.e., affine functions in the neural network).
The quadratic encoding of ReLU can be defined as follows.

\begin{definition}[QCQP encoding of ReLU] \label{def_reduction_qcqp}
Given a ReLU function $y=max\left(0,x\right)$, we can encode it to a set of quadratic constraints, as shown below:
$$
y=max\left(0,x\right) \Leftrightarrow \left( \left( y\ge x \right) \land  \left( y\ge 0 \right) \land \left(y \left(y-x\right)=0\right)\right)
$$
\end{definition}

Compared with LP and MILP encodings, the QCQP encoding enables us to analyze quadratic relations of symbolic (i.e., input and output of nodes) within a neural network model.
By applying the convex encoding for ReLU activation, the QCQP can be solved with soundness and completeness guarantee using off-the-shelf interior point methods.
The major concern with this type of encoding comes from scalability. Interior point methods are shown to be computationally expensive and could take up to an $O(n^6)$ complexity for a neural network model containing $n$ hidden units~\cite{dathathri2020enabling}. 

Existing papers that adopt QCQP encoding commonly mitigate the scalability issue in two directions. They apply various mathematical optimization to improve solving efficiency and/or take advantage of the linear approximation to further relax the original QCQP.
Particularly, recent work exploits linear approximation to extend the verification with a broader range of activation functions. Through the linear approximation (to be detailed in Section~\ref{sec:linear_approximation}), both tanh and sigmoid functions can be approximately formulated as a set of quadratic equations.

\subsection{Layerwise Propagation}

The layerwise propagation faithfully and precisely simulates the computation of a neural network model to solve the verification as a reachability problem. It is intuitive and has been adopted in some early approaches~\cite{scheibler2015towards,xiang2017reachable}.

Given the property specification, we first identify the domain of concerned input (i.e., all the possible adversarial inputs). Then we simulate the calculation based on those inputs until reaching the output layer. 
In the end, we compare the propagated outputs with the domain of expected output to determine if the property preserves in the target model. 
We visualize the layerwise propagation process of verifying the sample model in Fig.~\ref{fig:three-reduction}-(a).

The layerwise propagation at the reduction stage guarantees both soundness and completeness as it in fact does not explicitly alter the neural network as the body to be verified. 
The overall completeness and soundness are fully determined by the following reasoning strategies, which we will discuss in Section~\ref{sec:reasoning}.
Layerwise propagation usually works well on deep models, however, suffers while dealing with thick ones (i.e., lots of neurons per layer). 
Thus, the reasoning based on the layerwise propagation reduction without any optimization or trick would be computationally expensive, therefore leaving concern of scalability.

\begin{figure}[tb!]
	\begin{footnotesize}
		\vspace{-8pt}
		\begin{alignat*}{2}
			\text{minimize\,\,\,\,}   & x_{9}-2\times x_{10} + 1.5\ & \  \\
			\text{s.t.\,\,\,\,} &  x_1+x_2\le x_5\le x_1+x_2+M\cdot d_3, &\  \\[-2pt]
			& x_5\le M\cdot \left( 1-d_3 \right), \ &\ \\[-1.5pt]
			& x_1-x_2-1\le x_6\le x_1-x_2-1+M\cdot d_4, &\  \\[-1.5pt]
			& x_6\le M\cdot \left( 1-d_4 \right), \ &\  \\[-1.5pt]
			& x_5+x_6-1\le x_9\le x_5+x_6-1+M\cdot d_7, &\  \\[-1.5pt]
			& x_9\le M\cdot \left( 1-d_7 \right), \ &\  \\[-1.5pt]
			& x_5-x_6-1\le  x_{10}\le x_5-x_6-1+M\cdot d_8, &\  \\[-1.5pt]
			& x_{10}\le M\cdot \left( 1-d_8 \right), \ &\  \\[-1.5pt]
			& -1\le x_1\le 1,\ -1\le x_2\le 1, \ &  \\[-1.5pt]
			& d_3, d_4, d_7, d_8\in\left\{0,1\right\},\ x_5, x_6, x_9, x_{10} \ge 0,\ & \\[-1.5pt]
			& M > x_i, &\, \left(1\le i \le 12\right) 
		\end{alignat*}
	\end{footnotesize}
	\vspace{-20pt}
	\caption{MILP encoding of the sample network after equations simplifying, with the objective expected to be positive}
	\label{fig:milp-encoding}
\end{figure}

\subsection{Discretization}

\begin{figure*}[!t]
	\centering
	% \vspace{5pt}
	% Trim is performed in the order of {<left> <lower> <right> <upper>}
	% \includegraphics[trim=0cm 1.9cm 0cm 2.45cm, clip, width=0.88\linewidth]{images/three_types_of_reduction.pdf}
	\includegraphics[trim=0cm 1.9cm 0cm 7.75cm, clip, width=0.95\linewidth]{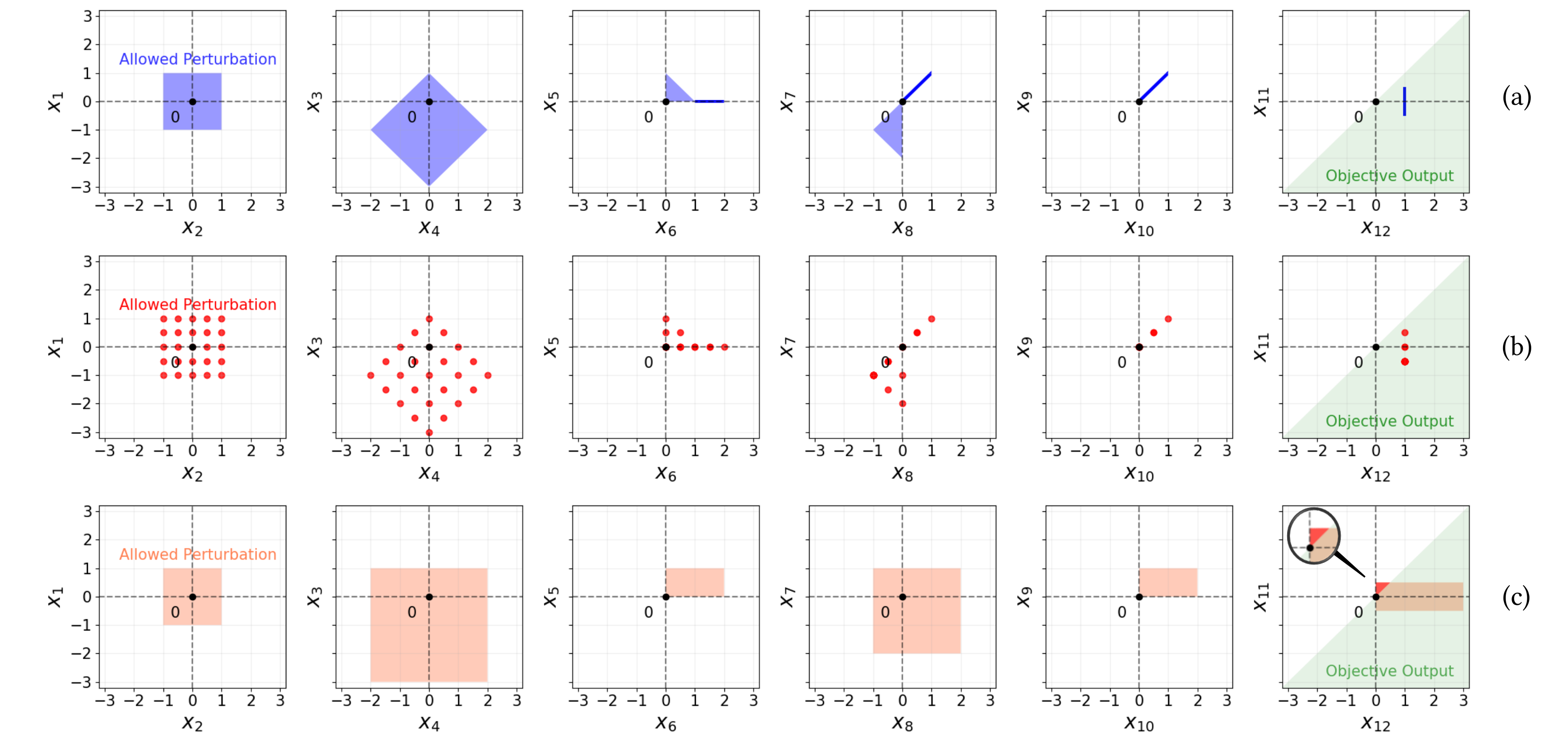}
	\vspace{-8pt}
	\caption{Visualization of different reduction approaches in verifying robustness property of the sample model (abstraction interpretation with box domain and reasoned by interval arithmetic in sub-figure \textbf{(a)}; discretization in sub-figure \textbf{(b)} and forward reachability analysis in sub-figure \textbf{(c)}, with red color area indicating false positive produced)}
	\label{fig:three-reduction}
\end{figure*}

The concern of scalability of layerwise propagation exists since the early stage of neural network verification. One possible approach to address such a concern is to mitigate the computational burden by \emph{discretization}.
Discretization is a concept in applied mathematics that describes the process of converting a continuous function into several discrete pieces, which aims to achieve a finite exploration within an infinite search space.
The discretization has been adopted in a verification approach named \textsc{DLV}~\cite{huang2017safety}, which proposes to discretize the perturbation from the infinite input distribution.
It propagates those discrete input vectors layer by layer and determines whether the robustness is preserved or not by observing the bounds of output. 
Fig.~\ref{fig:three-reduction}-(b) demonstrates the verification of the sample model through a discretization.

\begin{comment}
\begin{definition}[Discretization] \label{def_discret}
	Given a neural network model $f$ that takes input from a continuous domain $X$ and outputs to another continuous domain $Y$. The model $f$ can be discretized to $f_{d}$ with the entire network remains unchanged but has a different finite input and output domains as below:  
	
	$$
	\mathcal{M}_{\text{(Disc.)}} = f_{d}:X'\rightarrow Y'
	$$
	where $X'$ is the sampling set discretized from the continuous domain $X$, alternatively written as $X'\in X$. $Y'$ is the output corresponding to $X'$, therefore $Y'\in Y$.   
\end{definition}
\end{comment}

Although the architecture of the examined model remains unchanged, the domain of output completely depends on the discretized input domain. Unless the scale of input discretization has been set large enough to sufficiently depicts all the key points in the output domain, such as the \emph{convex hull} of a hidden layer, there always exists a risk of precision loss and consequently leads to missed violation (false negative). 
Overall, discretization greatly simplifies the analysis, especially for a deep neural network that takes comparably lower dimension inputs. 
However, as a reduction process, it does not come with a guarantee of completeness. 
The guarantee of soundness is challenging, too, as it is mutually restricted by the granularity of discretization and the convexity of the computation at each layer. 
However, many neural networks, particularly those that use sigmoid or tanh activation, are not convex. 
For these reasons, this reduction approach may suffer from the state-space explosion issue and therefore does not scale to larger neural network models~\cite{gehr2018ai2,wang2018formal}.

\subsection{Abstract Interpretation}

Abstract interpretation is a verification approach that provides a formalism of approximation and abstraction in a mathematical setting, which has been adopted in~\cite{gehr2018ai2,singh2018fast,yisrael2019abstraction} and their extensional works.
An abstract interpretation consists of an \emph{abstract domain}, a pair of \emph{abstraction and concretization functions}, and a sound \emph{abstract semantic function} in form of an iterative fixed-point equation. 

Abstract interpretation aims to simplify the verification problem at the reduction stage. Although the verification is still treated as a reachability problem, it is solved by reasoning an abstract system rather than the original neural network model. 
The mainstream abstraction interpretation approaches symbolize each node of the neural network, apply abstraction function over its computation, and ultimately determine whether the property is preserved through the output values after the concretization. Besides that, \cite{yisrael2019abstraction} explores another direction of the abstract interpretation that simplifies the architecture of the examined model, rather than approximating the output of neurons inside.

In \cite{yisrael2019abstraction}, each hidden layer has been simplified to contain at most four fully-connected neurons, literally one neuron with the maximum positive weight and tends to bring increment to the output written as $a^{[+,\uparrow]}$, one neuron with the minimize negative weight and tends to bring increment to the output $a^{[-,\uparrow]}$, one neuron with the maximum positive weight and tends to bring decrement to the output $a^{[+,\downarrow]}$, and lastly one neuron with the minimum negative weight and tends to bring decrement to the output $a^{[-,\downarrow]}$. Thus, both upper and lower bounds of the output can be eventually obtained by analyzing the abstract model. 
This approach gains a scalability advantage in verifying neural networks that are not very deep but built with a great number of hidden units per layer.
On the other hand, simplifying neural network architecture through clustering the neurons' output does not work well on models that adopt convergence functions as activation, such as sigmoid and tanh (because they always produce output within a certain range).

At the time of writing, \textsc{DeepPoly}~\cite{singh2019abstract} and its extensional approaches are the most comprehensive, precise, and practical solutions to perform verification by abstract interpretation. 
In the scheme of \textsc{DeepPoly}, each node is transformed into an abstract element. An abstract element is a tuple constituted by two \emph{polyhedral constraints} and two \emph{associate constants}. The polyhedral constraints of a node are made up of variables corresponding to those nodes connected from the previous layer, and the associate constants define its approximated upper and lower bounds after the concretization. 
% After the abstraction, the verification of a deep neural network is converted to the interpretation of abstract elements after being concretized. 
Take the sample network as an instance, the robustness verification is reduced to determine if, in its concretized abstraction, the upper bound of $x_{12}$ is less than the lower bound of $x_{11}$, which means $x_{11}>x_{12}$ is always true.
We take~\cite{singh2019abstract} as an example and define the abstract interpretation reduction as follows. 

\begin{definition}[Abstract interpretation] \label{def_abst_interp}
	Given a neural network model $f$ with $n$ nodes. 
	By defining a polyhedral abstract domain and approximating all the nodes accordingly, the verification of $f$ can be achieved by concretizing its abstraction $f_{a}$, written as:
	
	$$
	\mathcal{M}_{\text{(Abst.)}}=f_{a}:\left\{\forall i \in [n]\ .\ a_i=\left<a^{\le}_{i}, a^{\ge}_{i}, u_i, l_i\right>\right\} 
	$$
	where $a^{\le}_{i}$ and $a^{\ge}_{i}$ are polyhedral constraints of the abstract element corresponding to the $i$-th node, and $u_i$ and $l_i$ are associated constants that reflect the concretization of the $i$-th node's abstraction.   
\end{definition}

In Fig.~\ref{fig:three-reduction}-(c), we demonstrate the verification of the sample model by applying a box domain (each node is abstracted to a pair of lower and upper bounds) in the abstract interpretation. 
As we can observe from the output layer in the figure, the verification fails to determine whether the property holds due to a too coarse approximation (the tiny red area of the output domain \emph{violates} the property specification). That shows the given property specification could not be verified in the sample model through this reduction approach, although it holds in the reality.
Throughout the history of applying abstract interpretation in neural network verification, the abstract domain and corresponding abstraction functions have experienced evolution towards a more precise approximation of the neural network's output, aiming to enhance the effectiveness of the verification through this reduction. 
Fortunately, the same property can be successfully verified by later works such as \textsc{DeepPoly}~\cite{singh2019abstract}, which we will discuss in the next section and visualize in Fig.~\ref{fig:refinement}.

Overall, abstract interpretation reduction enables analyzing the examined model's behaviors in a less precise but more efficient way.
Compared with the discretization that performs analysis merely based on discrete values of the input, abstract interpretation has an advantage in soundness guarantee as it ensures that all possible values of a given node are within the bounds obtained from over-approximation.

\section{Specification Reasoning}
\label{sec:reasoning}
Specification reasoning discusses in detail which tool or algorithm to be adopted in performing the verification. 
The choice of reasoning strategy closely relates to the reduction approach applied to the examined model and determines the overall verification performance. In this section, we classify available reasoning strategies into five classes and present the result in columns 15-19 of Table~\ref{tab:table-all}. 
With various reduction approaches defined in advance, we establish the connection between each reasoning method and its applicable reduction setting. 
Here we remark that one verification approach may adopt multiple specification reasoning strategies.
Next, we detail each reasoning strategy and discuss its advantages and shortcomings. 

\subsection{Constraint Reasoning}

Constraint reasoning is an analysis approach implemented by an off-the-shelf solver, and therefore it is also known as \emph{solver reasoning}. This reasoning option is designed to take a set of constraints as input but does not restrict any reduction setting applied in advance.  
Reasoning through a satisfiability modulo theories (SMT) solver is the most intuitive way to perform verification of a neural network in SAT encoding. 
An SMT solver takes the SAT encoding of the examined model as input and heuristically searches for the best strategy to find a solution that satisfies all constraints. 

Using an SMT solver to verify a large and deep neural network containing hundreds or even thousands of neurons is not an efficient approach. With the worst case of \emph{exponential time complexity}, reasoning by an SMT solver usually cannot guarantee to terminate within a reasonable time. 
The high \emph{cyclomatic complexity} constitutes another challenge. 
As each ReLU node generates a disjunction due to its piecewise-linear nature, reasoning a deep neural network with $n$ ReLU neurons in hidden layers could produce $2^n$ sub-problems. 
Due to the lack of a polynomial-time algorithm for solving a CNF formula, constraint reasoning faces a critical challenge of scalability. Dedicated tools using SMT solvers to solve CNF encoding are shown only able to handle  small models containing 10-20 neurons in hidden layers~\cite{pulina2012challenging}.

One mitigation is replacing the SMT solver with an LP solver, which offers efficiency improvement in solving although sharing the same worst case of time complexity. For example, \textsc{Reluplex}~\cite{katz2017reluplex} adopts an LP solver with an in-house implementation of the \emph{simplex} algorithm in finding the feasibility of constraints. 
The simplex algorithm is known to be \emph{nondeterministic polynomial (NP)} in time complexity but a remarkably effective way to solve an LP problem by a series of \emph{update} and \emph{pivot} operations.
In \textsc{Reluplex}, a ReLU function is treated as a special equation. The value of ReLU output will be checked in each update operation.
To avoid the infinite pivoting on variables within one ReLU function (either its input or output), \textsc{Reluplex} also adopts a branch-and-conquer strategy to split the original problem into two sub-problems, with each sub-problem corresponding to the \emph{active} or \emph{inactive} mode of that ReLU node. 
The experiment of~\cite{katz2017reluplex} demonstrates a successful verification of 10 properties on the ACAS Xu\footnote{ACAS stands for Airborne Collision Avoidance System, where ACAS Xu is an experimental variant designed for the Remotely Piloted Aircraft Systems (RPAS).} system.
Owe to splitting heuristics, \textsc{Reluplex} could successfully verify neural network models with up to 300 ReLU nodes.
It for the first time marks robustness verification \emph{practical} in actual applications of deep neural networks. 
The latest progress of SMT/LP solver-based verification is~\textsc{Marabou}~\cite{katz2019marabou}, which brings a significant improvement over its former release \textsc{Reluplex} by introducing a symbolic tightening algorithm and integrating an improved simplex algorithm.

Using a solver over the MILP encoding is another mitigation to address the scalability concern because it resolves the ``branch explosion'' issue in the SAT/LP encoding. 
\emph{Gurobi} is the most used solver to reason the MILP encoding (see Table~\ref{tab:table-solver-reasoning}). Through the survey, we find~\cite{lomuscio2017approach,cheng2017maximum} might be the genesis of encoding a neural network into a MILP problem and reason by a solver. Later endeavors such as~\cite{dutta2018output,fischetti2019deep,tjeng2019evaluating} also make use of this set of encoding and reasoning strategies to apply different solvers and attempt to support more types of neural network models. 
According to the benchmarking performed in~\cite{dutta2018output}, reasoning property specification as a MILP problem brings significant improvement in time performance. It allows us to analyze larger models containing up to 6,000 neurons. 

Reasoning through a solver comes with an advantage that natively guarantees soundness and completeness, based on the premise that the solver is sound and complete. However, it faces scalability issues to a certain degree. 
Considering the desirable sound and complete verification can only be achieved if both modeling and reasoning approaches are sound and complete -- at this moment, a solver-based verification, especially the MILP solver-based verification, is still the best choice to verify a deep neural network with a guarantee of completeness.

\begin{table}
	\centering
	\caption{\label{tab:table-solver-reasoning} List of research works that disclose solver usage for the reasoning}
	\vspace{-8pt}
	% Config the row spacing
	\def\arraystretch{1.1}
	\begin{footnotesize}
	% Config the column spacing
	\setlength{\tabcolsep}{4pt}
	\setcounter{rownumber}{0}
	\begin{tabular}{HlHl}
		\hline
		\multirow{1}{*}{\textbf{No.}} & \textbf{\makecell[l]{Authorship, Year \& Reference}} & \multicolumn{2}{l}{\textbf{Solver(s)}}  \\%\cline{3-4}
		\hline
		\refstepcounter{rownumber}\therownumber & Pulina \& Tacchella, 2010 (\textsc{NeVer})\cite{pulina2010abstraction} & SMT & HySAT \\
		\refstepcounter{rownumber}\therownumber & Scheibler \textit{et al.}, 2015~\cite{scheibler2015towards} & SMT & iSAT \\  
		%\refstepcounter{rownumber}\therownumber & Bastani \textit{et al.}, 2016~\cite{bastani2016measuring} & SMT & (\emph{Not Specified})  \\
		\refstepcounter{rownumber}\therownumber & Cheng \textit{et al.}, 2017 (\textsc{MaxResilience})~\cite{cheng2017maximum} & MILP & CPLEX \\
		\refstepcounter{rownumber}\therownumber & Ehler, 2017 (\textsc{Planet})~\cite{ehlers2017formal} & SMT, LP & miniSat, GLPK \\
		\refstepcounter{rownumber}\therownumber & Huang \textit{et al.}, 2017 (\textsc{DLV})~\cite{huang2017safety} & SMT & Z3 \\
		\refstepcounter{rownumber}\therownumber & Katz \textit{et al.}, 2017 (\textsc{Reluplex})~\cite{katz2017reluplex} & LP & GLPK \\
		\refstepcounter{rownumber}\therownumber & Lomuscio \& Maganti, 2017 (\textsc{NSVerify})~\cite{lomuscio2017approach} & MILP & Gurobi \\
		\refstepcounter{rownumber}\therownumber & Dutta \textit{et al.}, 2018 (\textsc{Sherlock})~\cite{dutta2018output} & MILP & Gurobi \\
		% \refstepcounter{rownumber}\therownumber & Fischetti \& Jo, 2018~\cite{fischetti2019deep} & MILP & CPLEX \\
		\refstepcounter{rownumber}\therownumber & Wang \textit{et al.}, 2018 (\textsc{Neurify})~\cite{wang2018efficient} & MILP & Gurobi \\
		\refstepcounter{rownumber}\therownumber & Weng \textit{et al.}, 2018 (\textsc{Fast-Lin})~\cite{weng2018towards} & MILP & Gurobi \\
		\refstepcounter{rownumber}\therownumber & Bunel \textit{et al.}, 2018 (\textsc{BaB}, etc.)~\cite{bunel2018unified} & MILP & Gurobi \\
		\refstepcounter{rownumber}\therownumber & Katz \textit{et al.}, 2019 (\textsc{Marabou})~\cite{katz2019marabou} & LP & GLPK \\
		\refstepcounter{rownumber}\therownumber & Tjeng \textit{et al.}, 2019 (\textsc{MIPVerify})~\cite{tjeng2019evaluating} & MILP & Cbc, CPLEX, Gurobi \\
		\refstepcounter{rownumber}\therownumber & Singh \textit{et al.}, 2019 (\textsc{RefineZono})~\cite{singh2019boosting} & MILP & Gurobi \\
		\refstepcounter{rownumber}\therownumber & Singh \textit{et al.}, 2019 (\textsc{kPoly})~\cite{singh2019beyond} & MILP & Gurobi\\
		\refstepcounter{rownumber}\therownumber & Botoeva \textit{et al.}, 2020 (\textsc{Venus})~\cite{botoeva2020efficient} & MILP & Gurobi \\
		\refstepcounter{rownumber}\therownumber & Bak \textit{et al.}, 2020 (\textsc{NNenum})~\cite{bak2020improved} & LP & GLPK \\
		\refstepcounter{rownumber}\therownumber & Henriksen \& Lomuscio, 2020 (\textsc{VeriNet})~\cite{henriksen2020efficient} & MILP & Gurobi \\
		%\refstepcounter{rownumber}\therownumber & Kouvaros \& Lomuscio, 2021 (\textsc{Venus2})~\cite{kouvaros2021towards} & MILP & (\emph{Not Specified})\\
		\hline
	\end{tabular}	
	\end{footnotesize}
\begin{flushleft}
\textsuperscript{*} Approaches that claim their usage of solvers but do not disclose the names are not listed.
\end{flushleft}
\vspace{-15pt}
\end{table}

\subsection{Mathematical Optimization}

With the LP, MILP and SDP encoding selected, the verification of a neural network could also be achieved through various mathematical optimizations. 
Instead of generic programming (e.g., Python or C) with a specific constraint solver, approaches in this category are usually implemented based on off-the-shelf mathematical programming and numerical analysis platforms such as MATLAB. 

One typical technique in this type is \emph{duality}, which theoretically enables us to certify the robustness property by proving if the dual problem is solvable, although its original form is not practically easy to be solved. 
Wong and Kolter~\cite{wong2018provable} proposed the first work that falls in this class. It illustrates the distortion over benign input as an adversarial polytope and applies a MILP encoding for the examined model. 
After that, it encodes the property violation into the \emph{dual formulation} of the LP problem that corresponds to the original neural network model, and then applies convex \emph{outer approximation} (also known as \emph{relaxation}) in solving the dual problem. According to the duality theory, the original problem (also known as \emph{primal}) can be guaranteed with a lower bound of the solution if its dual problem is found feasible to be solved. 
Moreover, when the number of variables blasts with the growth of the size of neural networks, the dual formulation becomes comparably easier to be solved than its primal. 
As a result, it provides a novel direction to make the intractable verification problem feasible to be solved through mathematical optimization. 
% \zhe{you need to be careful when saying ``incomputable'' as it refers to a class of problems that are not computable/decidable, which means it's not possible to solve the problem at all in general, not to say solving them efficiently. I think the word you wanted to use was ``infeasible'' or ``intractable''.} -> Accepted, replaced with ''intractable''

Duality of LP/MILP only supports neural networks model with ReLU activation functions. Additionally, in case more non-piecewise-linear activation functions are supported or the examined neural network layers grow larger in size, the verification could suffer from false positives when the property fails to be certified -- which owes to the gap between the dual and primal problems.
 
There is another approach proposed by Raghunathan \textit{et al.}~\cite{raghunathan2018certified} that also adopts a similar idea of solving the dual of the original problem.
In that work, the encoding of neural networks is presented as an SDP that is composed of \emph{quadratic constraints}.
Besides that, it presents the dual problem in the form of \emph{Lagrange dual function} and uses gradient descent to solve the optimization problem. Since all mainstream activation functions are differentiable, this approach supports a wide range of activation functions, including sigmoid and ReLU. Nevertheless, that work does not show good scalability, too. It only performs well on a fully-connected MLP with up to two layers. 
% The conventional solving methods for SDP (e.g., interior point methods) could result in an $O\left(n^6\right)$ computational complexity~\cite{dathathri2020enabling} therefore it only performs well on a fully connected MLP with up to two-layers.
Recent work that applies a similar reasoning strategy to obtain the bound of the SDP problem includes~\cite{batten2021efficient,fazlyab2020safety,raghunathan2018semidefinite,dathathri2020enabling}.
Another work by Dvijotham \textit{et al.}~\cite{dvijotham2018dual} takes the limitation of previous research into account and proposes a general dual approach to support a broader class of neural networks. Compared with~\cite{wong2018provable,raghunathan2018certified}, it claims to obtain non-trivial robustness bounds without the need for bound optimization during adversarial training. However, the quality of the robustness bound is traded off for the enhancement in scalability. 

Reasoning by mathematical optimization is essentially introduced as an optimization technique in robust training. Unlike verifying a deep neural network model with a predefined property specification, this reasoning approach is more effective in finding the maximum robustness of a neural network model and helping in further neural network optimization. 
Reasoning by mathematical optimization faces challenges from both completeness and scalability. 
\cite{fazlyab2020safety} improves verification scalability, however, only for neural networks with up to five layers.
The disadvantage in scalability makes mathematical optimization less solely applied afterward.
A recent approach~\cite{batten2021efficient} integrates mathematical optimization with linear approximation (to be detailed in Section~\ref{sec:linear_approximation}) to enhance the overall efficiency, which presents a promising direction for future research.

\subsection{Interval Arithmetic}

Interval arithmetic, also known as interval analysis or bound propagation, is essentially developed as a classical method to measure the bounds of errors in mathematical computation~\cite{dawood2011theories}. 
It can also be used as an ideal method to evaluate the bounds of neural network output. 
To the extent of robustness specification reasoning, the preservation of robustness can be soundly certified once we establish the bounds of its output, which are propagated from the input perturbations, are within the acceptable range (i.e., the examined model does not misbehave).

Interval arithmetic has been adopted in early works~\cite{cheng2017maximum,ehlers2017formal} as scalability mitigation of SAT/LP solver-based verification. By symbolizing all nodes in the examined neural network model, each node can be represented as a composition of variables at its previous layer. 
Given predefined bounds of the input, the output of the neural network can be evaluated through a layer-by-layer analysis. Considering both affine and ReLU functions are made up of simple arithmetic operations, we can follow the basic rules of interval operation to generate bounds for them~\cite{alefeld2000interval}. 

Let $f:\mathbb{R}^{[J]}\rightarrow \mathbb{R}$ be the affine function of a neuron within hidden layers that executes the assignment $x_i\gets b+\sum_{j\in \left[ J \right]}{\left(w_j\cdot a_j\right)}$, where $x_i$ has $J$ neurons in the previous layer connected to it, and $a_j$ stands for the output of the $j$-th neuron on the previous layer. Each $a_j$ has a pair of the lower and upper bounds written as $a_{j}^{\le}$ and $a_{j}^{\ge}$. The interval of $x_i$, represented by $\left[x_{i}^{\le},x_{i}^{\ge}\right]$, can be obtained as follows:

$$
	x_{i}^{\le}=b+\sum_{j\in\left[J\right]}{\left(max(0,w_j)\cdot a_{j}^{\le} + min(0,w_j)\cdot a_{j}^{\ge}\right)}
$$
$$
	x_{i}^{\ge}=b+\sum_{j\in\left[J\right]}{\left(max(0,w_j)\cdot a_{j}^{\ge} + min(0,w_j)\cdot a_{j}^{\le}\right)}
$$

Similarly, let $f:\mathbb{R}\rightarrow \mathbb{R}$ be the ReLU activation function that executes the assignment $a_i\gets max(0,x_i)$, where $x_i$ is the output of an affine function ahead and the input of the activation function. The bounds of $a_i$ can be obtained as follows:

$$
	a_{i}^{\le}=max(0,x_{i}^{\le})
\,,\,\,
%$$
%$$
	a_{i}^{\ge}=max(0,x_{i}^{\ge})
$$

We remark that interval arithmetic does not take symbolic dependency into account so it tends to generate a larger interval for a deep neural network when the number of layers increases. 
This phenomenon is also reflected in Fig.~\ref{fig:three-reduction}-(c) that the interval obtained at the output layer spans over the boundary of the objective (shown as the red color region), thus it fails to verify the robustness preservation.
For that reason, even though interval arithmetic is simple and fast in computation, it can only provide a coarse evaluation of the examined neural network’s output. 

To mitigate the issue caused by too coarse evaluation, an iterative refinement (to be detailed in  Section~\ref{sec:iterative-refinement}) sometimes is necessary to maintain the minimum bounds of neuron output in practice. For example, we can keep the symbolization along with layerwise propagation and then simplify the symbolic composition through backtracking to find a more precise interval of each neuron. Counterexample-based refinement is another useful strategy to refine the interval. 

As a reasoning strategy, the usage of interval arithmetic is usually combined with linear approximation in recent literature as scalability becomes the primary issue in verifying larger and deeper neural networks. On the other hand, integrating interval arithmetic with a proper approximation method can also enable verification for those neural networks that use an activation without a piecewise-linear character like sigmoid and tanh functions.

\subsection{Linear Approximation}
\label{sec:linear_approximation}

In deep neural network verification, linear approximation (also known as linear relaxation) aims to provide an estimated range of non-linear activation output. It is firstly used as a complement to solver-based verification to extend its support with a broader range of activation functions~\cite{pulina2010abstraction,pulina2012challenging}. 
Linear approximation is one of the primary reasoning techniques for abstract interpretation modeling. By applying linear approximation with layerwise symbolic analysis, it is expected to get rid of reliance on SMT/MILP solvers and achieve a verification with provable soundness and scalability.

Next, we choose three typical activation functions that are used in most feedforward neural networks and present the existing linear approximation algorithms proposed in the literature. Some linear approximation algorithms also correspond to the abstraction function (e.g., zonotope~\cite{singh2018fast} and polyhedron~\cite{singh2019abstract}) from the abstract interpretation perspective. 

\begin{figure}[t!]
	\centering
	% Trim is performed in the order of {<left> <lower> <right> <upper>}
	\includegraphics[trim=0cm 0cm 0cm 0cm, clip, width=0.85\linewidth]{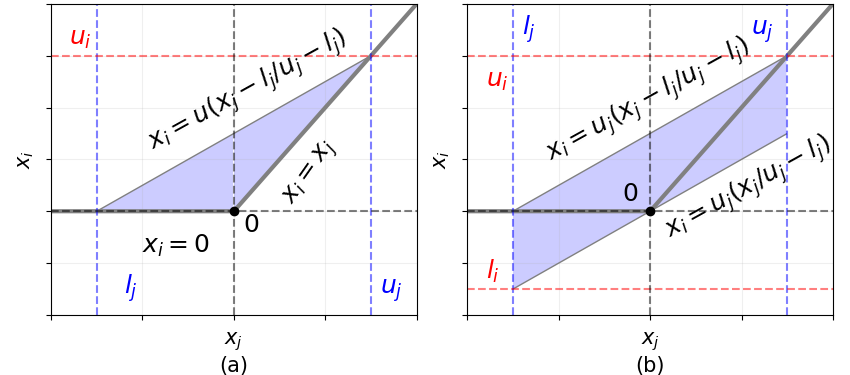}
	\vspace{-8pt}
	\caption{ReLU function and its convex approximation of the upper bound (\textbf{a}) and both the upper and lower bounds (\textbf{b}), the shaded regions represent the range of value approximation.}
	\label{fig:relu_various_approx}
\end{figure}
\begin{figure}[t!]
	\centering
	% Trim is performed in the order of {<left> <lower> <right> <upper>}
	\includegraphics[trim=0cm 0cm 0cm 0cm, clip, width=0.85\linewidth]{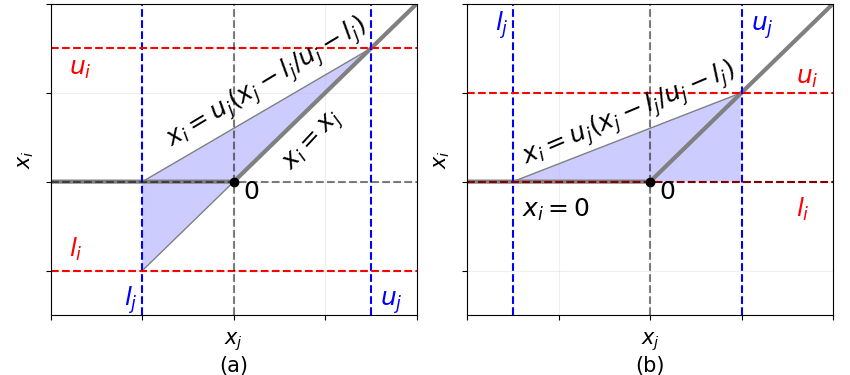}
	\vspace{-8pt}
	\caption{ReLU function and its convex approximation in \textsc{DeepPoly} for case (3) where $\lambda=0$ (\textbf{a}) and $\lambda=1$ (\textbf{b}), the shaded regions represent the range of value approximation.}
	\label{fig:relu_approx}
\end{figure}

\subsubsection*{\textbf{ReLU Function}}

Linear approximation of the ReLU function is firstly proposed in~\cite{ehlers2017formal} as a supplementary to the native LP modeling, which replaces the piecewise feature of the ReLU function by offering a pair of bounds, and thereby forms a constant mapping relation between the input and output of ReLU regardless of its activation status (i.e., active or inactive). Fig.~\ref{fig:relu_various_approx}-(a) illustrates an early linear approximation of ReLU with an upper bound. Another similar approximation that appears in~\cite{weng2018towards,wang2018efficient,zhang2018efficient} estimates both the upper and lower bounds of ReLU output in case the activation mode is unable to be determined (i.e., the input range spans over zero), as shown in Fig.~\ref{fig:relu_various_approx}-(b).

The state-of-the-art linear approximation of the ReLU function is proposed in \textsc{DeepPoly}~\cite{singh2019abstract} as the abstraction function over a polyhedral domain. 
Recall that in Definition~\ref{def_abst_interp} we propose a sound approximation of a node's value by two polyhedral constraints (i.e., $a^{\le},\,a^{\ge}$) and two associate constants (i.e., $l,\,u$). 
Let $f:\mathbb{R}\rightarrow \mathbb{R}$ be a function that executes the assignment $x_i\gets \max \left( 0,x_j \right) $ for $j=i-1$. 
Given $\left< a^{\le}_j,a^{\ge}_j,l_j,u_j \right>$,  we can estimate the output in three cases:

	\begin{enumerate}
		\item If $l_j\geqslant0$, $a_i^{'\le}\left( x \right) = a_i^{'\ge}\left( x \right) = x_j$, $l'_i=l_j$, and $u'_i=u_j$;\vspace{3pt}
		\item If $u_j\leqslant0$, $a_i^{'\le}\left( x \right) = a_i^{'\ge}\left( x \right) = 0$, $l'_i=u'_i=0$;\vspace{3pt}
		\item Otherwise $a_{i}^{'\le}\left( x \right) =\lambda \cdot x_j$, $a_{i}^{'\ge}\left( x \right) =u_j\frac{\left( x_j-l_j \right)}{\left( u_j-l_j \right)}$, $l'_i=\lambda \cdot x_j$, $u'_i=u_j$ where $\lambda \in \left\{ 0,1 \right\} $ that minimizes the area of the resulting shape in the ($x_i$,$x_j$)-plane. i.e., if $\left| l_j \right|\geqslant \left| u_j \right|$, then $\lambda=0$; otherwise $\lambda=1$. Fig.\ref{fig:relu_approx} illustrates two possible scenarios of the approximation in this case.
	\end{enumerate}
	
Compared with previous literature, the approximation proposed in \textsc{DeepPoly} produces the smallest approximation region (shown as shadowed areas in Fig.~\ref{fig:relu_approx}), which stands for the highest precision among all relevant works.

\subsubsection*{\textbf{Sigmoid and Tanh Functions}}

Linear approximation of the sigmoid function is firstly proposed in~\cite{pulina2010abstraction} where the input and output of a sigmoid function ($\sigma$) are mapped with intervals. By doing that, a sigmoid function can be encoded into linear constraints containing input and output variables, and then be reasoned by an SMT/LP solver. 

\begin{figure}[t!]
	\centering
	% Trim is performed in the order of {<left> <lower> <right> <upper>}
	\includegraphics[trim=0cm 0cm 0cm 0cm, clip,width=0.85\linewidth]{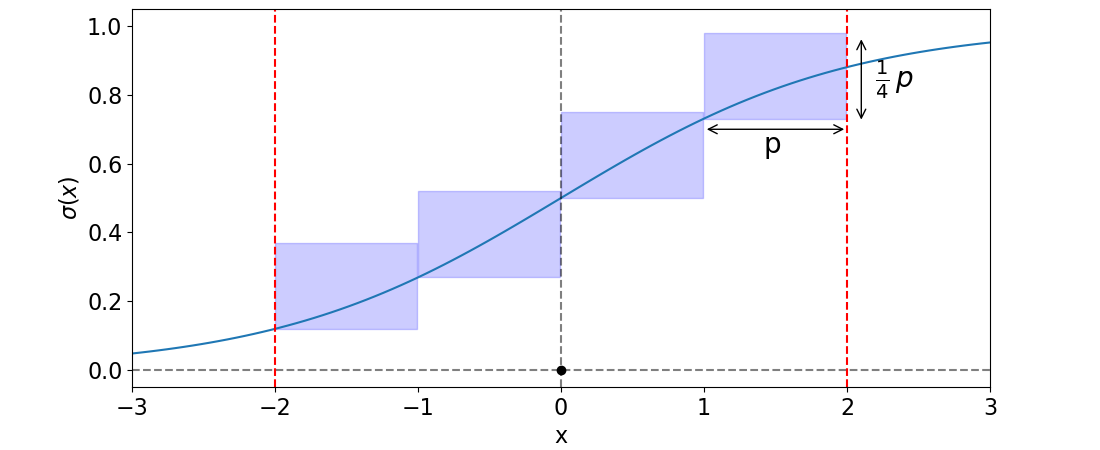}
	\vspace{-10pt}
	\caption{Interval abstraction proposed to evaluate the output of the sigmoid function, with input ranges over $[-2,2]$ and $p=0.5$.}
	\label{fig:pulina_sigmoid_interval_abst}
\end{figure}

The output range of a sigmoid function is fixed as $\left(0,1\right)$ and its derivation, $\sigma'(x) = \sigma(x)(1 - \sigma(x))$, has the maximum value of $1/4$ at which the input is zero. As shown in Fig.\ref{fig:pulina_sigmoid_interval_abst}, if we divide the input domain as a series of intervals in the length of $p$, we can always approximate the bounds of the sigmoid activation function with an arbitrary input $x$ as $0\le \sigma\left(x+p\right) - \sigma\left(x\right) \le p/4$. Thus we can initiate a sound abstract interpretation with a box abstract domain. 

A similar method could also be applied to the tanh activation function since both of them have the maximum derivation value at zero points (the maximum derivation of $tanh\left(x\right)$ is 1 where $x=0$). 
This idea closely relates to another linear approximation by calculating the \emph{Lipschitz continuous gradient} to assess output bounds, which becomes commonly adopted in later research such as~\cite{weng2018towards,huang2019reachnn,fazlyab2020safety}.

\begin{figure*}[t!]
	\centering
	% Trim is performed in the order of {<left> <lower> <right> <upper>}
	\includegraphics[trim=0cm 0.5cm 0 0cm, clip, width=0.97\linewidth]{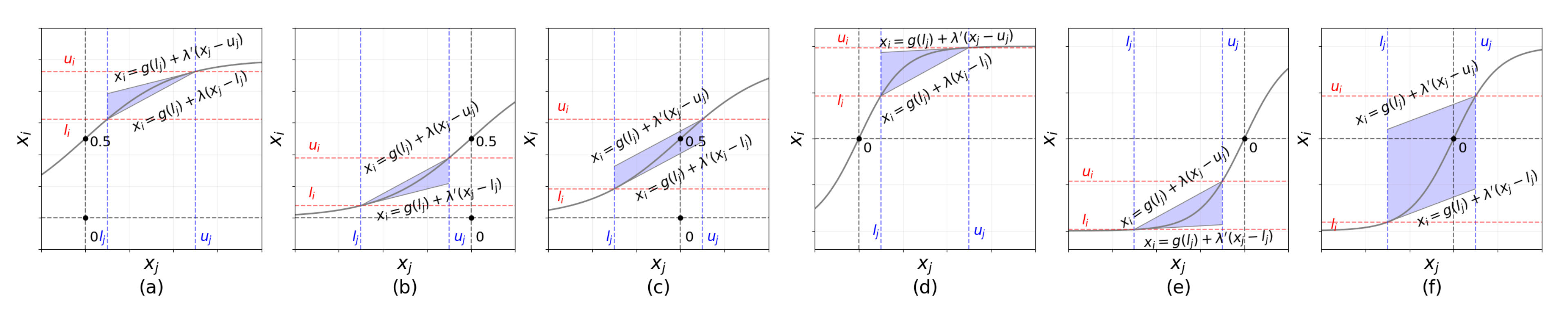}
	\vspace{-5pt}
	\caption{Visualization of the approximation of sigmoid/tanh activation that is proposed in \textsc{DeepZ} based on its transformation in case (2) where $l_j\geqslant0$ (\textbf{a} for sigmoid,\textbf{d} for tanh), $u_j<0$ (\textbf{b} for sigmoid,\textbf{e} for tanh), and $l_j<0 \leqslant u_j$ (\textbf{c} for sigmoid,\textbf{f} for tanh), the shaded regions represent the range of over-approximation.}
	\label{fig:sig_tanh_approx}
	\vspace{-5pt}
\end{figure*}

One recent linear approximation technique for sigmoid and tanh functions is proposed in \textsc{DeepZ}~\cite{singh2018fast} as a part of zonotope abstract interpretation and later adopted in its succeeding works, such as \textsc{DeepPoly}\cite{singh2019abstract}. 
Since both sigmoid and tanh functions present an S-shape on the input-output plane, they are always twice differentiable. Moreover, we observe their derivatives keep growing with the input value in the negative range, and conversely, declining with input's growth when the input becomes greater than zero. 
According to that character, we can estimate the growing slope of the output within a closed range determined by the lower and upper bounds of the input. 
This approximation is visualized in Fig.~\ref{fig:sig_tanh_approx} and detailed as follows.

\begin{figure}[t!]
	\centering
	% Trim is performed in the order of {<left> <lower> <right> <upper>}
	\includegraphics[trim=0cm 0cm 0 0cm, clip, width=0.83\linewidth]{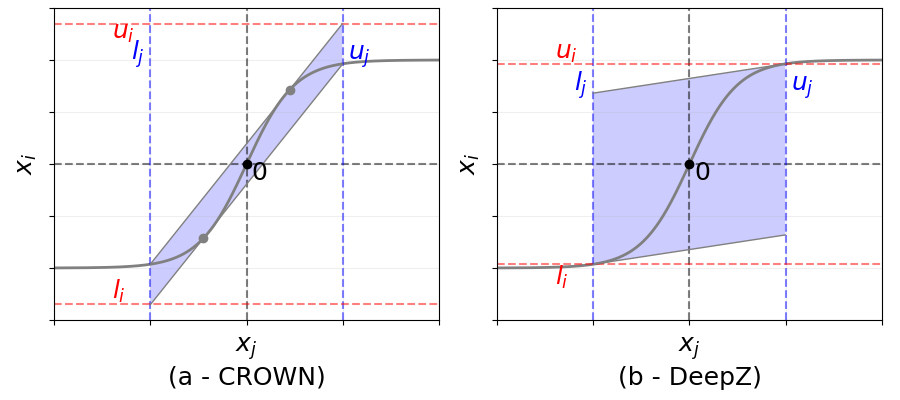}
	\vspace{-10pt}
	\caption{The convex approximation of the tanh function proposed in \textsc{CROWN} (\textbf{a}) and \textsc{DeepZ} (\textbf{b}) when $l_j<0 \leqslant u_j$, the shaded regions represent the range of over-approximation.}
	\label{fig:tanh_crown}
\end{figure}

Let $g:\mathbb{R}\rightarrow \mathbb{R}$ be a function that performs $x_i\gets \sigma\left( x_j \right) $ or $x_i\gets tanh\left( x_j \right) $ for $j=i-1$. 
Given $\left< a^{\le}_j,a^{\ge}_j,l_j,u_j \right>$, we can approximate the new component $i$ by setting two associate constants $l_i=g(l_j)$, $u_i=g(u_j)$, and two polyhedral constraints $a^{\le}_i$ and $a^{\ge}_i$ according to two cases below:
	\begin{enumerate}
		\item If $u_j=l_j$, then $a_i^{'\le}\left( x \right) = a_i^{'\ge}\left( x \right) = g(l_j)$; \vspace{3pt}
		\item Otherwise, we compute $a_i^{'\le}\left( x \right)$ and $a_i^{'\ge}\left( x \right)$ separately. \vspace{3pt}\\
		We first set $\lambda =\frac{g\left( u_j \right) -g\left( l_j \right)}{u_{j_{}}-l_j}$,\vspace{3pt}\\
		\hspace*{30px}and $\lambda '=\min \left( g'\left( u_j \right) ,g'\left( l_j \right) \right) $, then:\\
		(a) For the lower bound polyhedral constraint $a_i^{'\le}\left( x \right)$, we have:
				$$
				a_i^{'\le}\left( x \right) = \begin{cases}
					g\left( l_j \right) +\lambda \left( x_j-l_j \right) &		\text{if}\,l_j\geqslant 0\\
					g\left( l_j \right) +\lambda' \left( x_j-l_j \right) &		\text{if}\,l_j < 0\\
				\end{cases}
				$$
			\\(b) For the upper bound polyhedral constraint $a_i^{'\ge}\left( x \right)$, we have:
				$$
				a_i^{'\ge}\left( x \right) = \begin{cases}
				g\left( u_j \right) +\lambda \left( x_j-u_j \right) &		\text{if}\,u_j\leqslant 0\\
				g\left( u_j \right) +\lambda '\left( x_j-u_j \right) &		\text{if}\,u_j > 0\\
				\end{cases}
				$$
	\end{enumerate}

Another approximation is introduced in \textsc{CROWN} framework~\cite{zhang2018efficient} and applied in later research such as~\cite{henriksen2020efficient}. It proposes a similar approximation of sigmoid and tanh functions with \textsc{DeepZ} but offers more precise bounds when the input range spans over zero.
Considering sigmoid and tanh functions always be \emph{convex} given the input is less than zero, and be \emph{concave} for the positive input, \textsc{CROWN} estimates the upper and lower bounds of sigmoid or tanh function at an arbitrary input point through a \emph{tangent line} that originates from the input bounds, and thereby constructs a quadrilateral boundary. 

Here we take the tanh function (written as $g$) as an example. Suppose the upper and lower bounds of input are given as $[l_j,u_j]$ such that $l_j<0<u_j$, we call the curve segment corresponding to the input bounds $g_{[l,u]}$. There must exist a tangent line starting from either the lower bound or the upper bound of the tanh function touching the curve itself.
As shown in Fig.~\ref{fig:tanh_crown}(a), these two tangent lines constitute the upper and lower bounds of linear approximation. Since both tanh and sigmoid functions share the same shape of curve but only differentiate in output range, we omit the demonstration of the sigmoid function to save space. 

Notwithstanding finding the tangent line for each bound could be computationally costly, it offers a sound estimation with higher precision. \textsc{CROWN}~\cite{zhang2018efficient} proclaims the finding of tangent line can be achieved through a binary-search algorithm, which means a logarithmic time complexity on average and therefore theoretically does not escalate the overall difficulty of verification. 
Compared with the approximation proposed in \textsc{CROWN}~\cite{zhang2018efficient}, the zonotope abstraction of sigmoid and tanh functions proposed in \textsc{DeepZ}~\cite{singh2018fast} does not require finding tangent lines in calculating the output bounds. However, that approach produces a comparably coarse approximation (see Fig.~\ref{fig:tanh_crown}), where the slope at either the lowest input or the highest input tends to be zero -- in contrast \textsc{CROWN}~\cite{zhang2018efficient} does not have such concern. Overall, the linear approximation strategy of \textsc{DeepZ}~\cite{singh2018fast} has the potential to achieve a better time efficiency in the practice but sacrifices precision to a certain degree. 

\subsection{Iterative Refinement}
\label{sec:iterative-refinement}

Iterative refinement has been proposed as a complementary technique to reinforce incomplete analysis such as over-approximation. Some existing approaches~\cite{wang2018formal,singh2019abstract,bak2020improved,henriksen2020efficient} adopt it to achieve a \emph{complete} verification despite diverse incomplete reasoning strategies being used, while most of the others take advantage of it to mitigate the incompleteness.
It is firstly used in~\cite{pulina2010abstraction} to mitigate false positives in searching for the violation of a property specification. 
The iterative refinement strategy is inspired by the counterexample-guided abstraction refinement (CEGAR) in model checking. However, in the verification of neural networks, incomplete analysis usually does not offer a counterexample when the property specification is shown unsatisfied.
In that case, the goal of iterative refinement is to adjust the ``coarseness'' of the approximation for certain nodes within the examined model and eventually validate if the unsatisfied result is caused by a counterexample or it is merely a false alarm. 

Next, we categorize different techniques that have been used as iterative refinement by the existing approaches.

\begin{figure*}[t!]
	\centering
	% Trim is performed in the order of {<left> <lower> <right> <upper>}
	\includegraphics[trim=2cm 10.5cm 2cm 2.4cm, clip, width=1\linewidth]{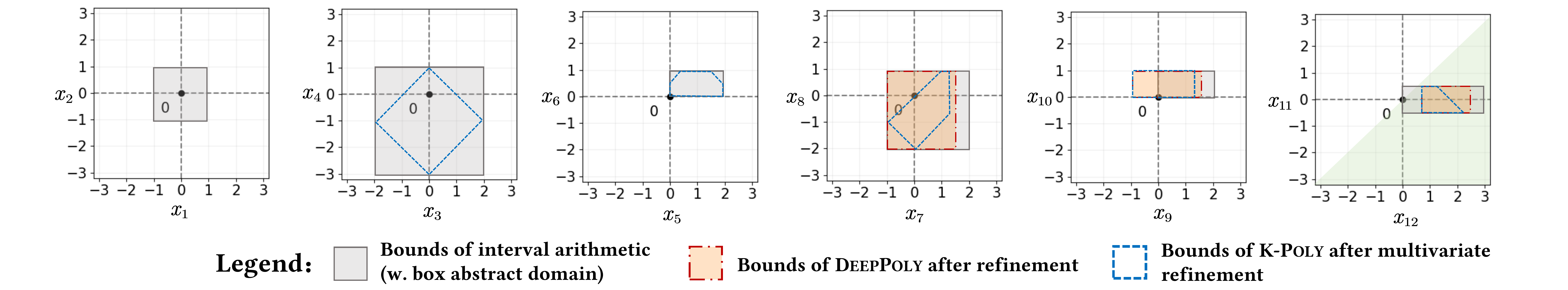}
	\vspace{-13pt}
	\caption{Visualization of layerwise bounds refinement over the abstract interpretation of the sample neural network (finer bounds in orange color are obtained through interval arithmetic based on polyhedral abstract domain aided with backtracking in \textsc{DeepPoly}, and bounds in blue colors are further refined by applying multiple variate analysis that is adopted in \textsc{kPoly})}
	\label{fig:refinement}
	\vspace{-8pt}
\end{figure*}

\subsubsection*{\textbf{Constraint Replenishment}}
Constraint replenishment is widely-used when there is no parameter available to improve the precision, and therefore we have to bring more constraints for a less coarse analysis. 
A common way to achieve that is by adding a complete reasoning technique, for example referring the MILP encoding to the ReLU function to introduce additional constraints over its output, making use of these additional constraints in the feedforward analysis, and in the end obtaining a less coarse approximation of the neural network output~\cite{wang2018efficient,singh2019boosting}.
Recent literature such as~\cite{singh2019boosting,kouvaros2021towards,bak2020improved,botoeva2020efficient,henriksen2020efficient} mainly perform linear approximation in the beginning, especially for the purpose of relaxing non-linear activation in a linear format, and then apply an off-the-shelf cross-platform solver library (e.g., Gurobi) to execute the verification. 

\subsubsection*{\textbf{Branch and Bound}}
Branch and bound (colloquially known as B\&B, BaB) is an algorithm design paradigm for solving NP-hard optimization and interval analysis~\cite{clausen1999branch}. 
It is in nature a linear relaxation of non-linear computation for constraint reasoning.
The branching takes advantage of an efficient \emph{searching} strategy to prioritize the ReLU nodes for the analysis, followed by the bounding process that exploits the cutting-edge over-approximation techniques to find the bounds of the node.

Branch-and-bound is firstly adopted in~\cite{pulina2010abstraction} as a parameter justification technique to exclude unsatisfied verification outcomes caused by spurious counterexamples. Since the coarseness of the sigmoid approximation is determined by the size of an interval of $p$ (see Fig.~\ref{fig:pulina_sigmoid_interval_abst}), higher precision can be achieved by heuristically assigning a smaller value to $p$. 
It repeats until we can determine if the property specification is satisfied, otherwise a counterexample that violates the property specification is given. 

A similar strategy has also been applied in approximating the ReLU activation in ~\cite{katz2017reluplex,ehlers2017formal,bunel2018unified,bak2020improved,batten2021efficient} by bisecting its input domain.  
\textsc{Reluplex}~\cite{katz2017reluplex} applies this strategy in their simplex algorithm to split each ReLU constraint into two constraints, i.e., the negative part that always leads to a zero output, and the positive part that remains unchanged through the ReLU computation.
In~\cite{bunel2018unified}, branch-and-bound has been adopted with multiple reasoning options to achieve completeness to different degrees. \textsc{NNenum}~\cite{bak2020improved} implements an enumerated searching algorithm in finding the sign of the ReLU input to refine bounds obtained from over-approximation and claims to achieve a complete verification in the end.
Another recent work~\cite{batten2021efficient} also takes advantage of this approach to aid constraint reasoning, performing 50\% better than the top candidates in recent verification competitions~\cite{vnncomp2020}. 

\subsubsection*{\textbf{Multiple Neurons Analysis}}
Multiple neurons analysis is an emerging strategy that is proposed to enhance precision and mitigate incompleteness. It refines the spaces of each symbolic through a multiple variate analysis, in an intra-layer or cross-layer manner.
It is firstly proposed in \textsc{kPoly}~\cite{singh2019beyond} called \emph{k-polyhedral convex hull approximation}, which reasons not only the constraints of feedforward propagation but also the linear relations of multiple neurons in the same layer. 
In addition to its previous work~\cite{singh2019abstract}, \textsc{kPoly} (by setting $k=2$) calculates the bounds of sum ($x_i + x_j$) and difference ($x_i - x_j$) of two nodes at each layer.
As shown in Fig.~\ref{fig:refinement}, by analyzing two nodes simultaneously, we can achieve higher precision than other approaches with only a single neuron analysis~\cite{singh2019abstract} along with the layer propagation. 
As a result, the size of bounding boxes at the output layer drops exponentially with the number of neurons to be analyzed at the same layer, implying less unnecessary precision loss produced due to the abstract interpretation. 

Another recent paper~\cite{tjandraatmadja2020convex} uses a similar strategy but with different reduction and reasoning strategies. It applies multivariate space refinement in both layerwise propagation (\textsc{FastC2V}) and LP solving (\textsc{OptC2V}), and thereby achieves better completeness than previous approaches at the price of sacrifice in scalability. 

Compared with other iterative refinement approaches, nowadays multiple neurons analysis still heavily relies on specific mathematical libraries or solving tools and therefore faces challenges to be adopted in generic programming or well-known machine learning frameworks (e.g., PyTorch, TensorFlow, etc.). 

\section{Future Directions}
\label{sec:discussion}
The upsurge of research in user-guide neural network verification started since the imperceptible input perturbation is proved to cause misclassification in several research works~\cite{szegedy2013intriguing,goodfellow2015explaining} and more importantly, in real-life applications~\cite{latimes2019tesla}. 
We find dozens of works published during the past decade that attempt to address this challenge and verify that a neural network model is robust against adversarial input perturbations. 
Next, we discuss a few potential research directions that are potential to bring progress in the future.

\subsubsection*{\textbf{Endeavor for a better completeness}}
Our research community has been relentlessly working to strike a balance between completeness and scalability. To resolve this dilemma, one potential solution could be a heuristic combination of techniques used in both incomplete and complete verifiers. In one of the most recent works by Singh \textit{et al.}~\cite{singh2019boosting}, a methodology as a mixture of abstract interpretation and MILP solving is proposed, where two techniques jointly take effect during the verification. After the over-approximation is finished on each layer, a solver starts refining the boundary obtained from approximation with a timeout (because it is a complete verification method that does not guarantee a termination). That idea has preliminarily proved to gain an advantage from both complete and incomplete verifiers. However, due to the limitation of the state-of-the-art MILP approaches, only piecewise-linear activation (i.e., ReLU) is supported in that work, which means there is still much work to do in the future to implement a practical tool for the industry.

\subsubsection*{\textbf{Adoption of various complex models}}
Most of the existing studies only perform the verification on MLP models, which is an elementary neural network architecture from nowadays view. 
As the max-pooling function could be treated as a variant of the ReLU function that outputs the maximum from multiple values, we are pleased to see some recent studies (e.g.,~\cite{gehr2018ai2,wong2018provable,singh2019abstract,tran2020nnv}) have adopted their verification approaches to convolutional neural networks (CNNs). 
However, deep neural networks are not created only for image classification. Robustness is equally important for voice recognition and text prediction tasks, which are achieved by diverse complex models, such as recurrent neural networks (RNNs) and transformers. In that case, the challenges of soundness and completeness trade-off and support of diverse types of functions (not necessarily to be activation function) become critical and urgent.
Recent research has made effort on extending existing reduction and reasoning techniques to robustness verification of RNNs~\cite{du2020mbr}, graph neural networks (GNNs)~\cite{wang2021certified}, $L_{\infty}$ distance networks~\cite{zhang2021boosting}, variational auto-encoders (VAEs)~\cite{dathathri2020enabling}, and transformers~\cite{shi2019robustness}. 
We foresee the existing approaches will be one day applied to more types of complex models.

\subsubsection*{\textbf{Leveraging optimization algorithms}}
Existing LP/MILP encoding-based methods rely on mathematical optimization formulations and attempt to find the minimum value of an objective function with a large number of constraints. If the minimum value is positive, then the verification answers true. While the state-of-the-art is successful in exploring the optimization state-space using advanced branching and relaxation techniques~\cite{kouvaros2021towards,batten2021efficient}, there is still potential room for improvement. One possibility is to leverage recent advancements of nature-inspired optimization algorithms~\cite{mirjalili2020nature} at the reasoning stage. 
Such algorithms are highly efficient in finding optimal values of extremely complex objective functions at the cost that the results are often approximations rather than the global maximum/minimum.
In the setting of the verification problem, this corresponds to trading completeness for scalability. However, with a customized algorithm dedicated to verification, it is possible to compute very close approximations of the minimum value for a large deep neural network in a significantly shorter time than existing methods. The above can be integrated into a method that combines incomplete and complete algorithms and takes advantage of both worlds \`a la~\cite{singh2019boosting}. It is also worth exploring the possibility of addressing different activation functions using such optimization algorithms. The outcome of these directions should be valuable despite being incomplete. 

\subsubsection*{\textbf{A generic verification framework}}
The advancement of verification techniques cannot benefit real-life AI applications unless it can be implemented in generic programming frameworks and well supports mainstream machine learning ecosystems. 
We find that much literature in this field releases a proof-of-concept that can only work on particular models and mathematical programming platforms like MATLAB. Only around 30\% (13 out of \nofwork, see Table~\ref{tab:table-all}) of surveyed papers have implemented their approaches based on TensorFlow and/or PyTorch, which are the two most popular deep learning frameworks as of 2022~\cite{oconnor2021pytorch}.
This phenomenon seriously limits the industrial adoption of robustness verification. 
We advocate that efforts need to be made for a more generic verification framework. Future research is suggested to release their techniques as public packages that are compatible with the mainstream deep learning ecosystems.

\section{Conclusion}
\label{sec:conclusion}
The verification of neural networks is the art of problem-solving.
Its development engages persistent research efforts from verification, optimization, and machine learning communities.
In this survey, we collect \nofwork existing approaches to verifying adversarial robustness and we propose our taxonomy for classifying those related works from a perspective of formal verification.
We present our classification result regarding three key components of model checking including the formalization of property specification, reduction of the examined model, and available reasoning strategies. 
To gain an in-depth understanding of the rationality and difference between those diverse approaches, we create a sample multi-layered neural network model and provide substantial visualization in the paper.
We then discuss both advantages and challenges that are yet to be addressed for each approach. In the end, we share our outlook on the future directions of deep neural network verification.

\bibliographystyle{IEEEtran}
\bibliography{reference}

\end{document}